\documentclass[epsf,useAMS,usenatbib, usegraphicx]{mn2e}
\usepackage{graphicx}
\usepackage{epsfig}
\usepackage{amsmath}
\usepackage{amssymb}
\usepackage{rotating}
\usepackage{color}
\usepackage{pifont}
\usepackage{longtable}
\usepackage{pdflscape,lscape}
\usepackage{appendix}
\newcommand{\ion}[2]{{#1}\,{\sc #2}}
\newcommand{\teff}{\mbox{$T_{\rm eff}$}}
\newcommand{\logg}{\mbox{$\log g$}}
\newcommand{\vsini}{\mbox{$v \sin i$}}
\newcommand{\kms}{\mbox{km\,s$^{-1}$}}

\usepackage{url}

\title[Eclipsing binary stars with a $\delta$~Scuti component]{Eclipsing binary stars with a $\delta$~Scuti component}

\author[F. Kahraman Ali\c{c}avu\c{s} et al.]{F. Kahraman Ali\c{c}avu\c{s}$^{1,2}$\thanks{E-mail: filizkahraman01@gmail.com\,/\,filizkahraman@comu.edu.tr}, E. Soydugan$^{1,2}$, B. Smalley$^{3}$ and J. Kub\'at$^{4}$
\\
$^{1}$Faculty of Sciences and Arts, Physics Department,Canakkale Onsekiz Mart University, 17100, Canakkale, Turkey\\
$^{2}$Astrophysics Research Centre and Ulup{\i}nar Observatory, \c{C}anakkale Onsekiz Mart University, 17100 \c{C}anakkale, Turkey\\
$^{3}$Astrophysics Group, Keele University, Staffordshire ST5 5BG, UK \\
$^{4}$Astronomick\'{y} \'{u}stav, Akademie ved \v{C}esk\'{e} republiky, CZ-251 65 Ond\v{r}ejov, Czech Republic \\
}

\begin{document}

\date{Accepted ... Received ...; in original form ...}

\pagerange{\pageref{firstpage}--\pageref{lastpage}} \pubyear{2017}

\maketitle

\label{firstpage}

\begin{abstract}
Eclipsing binaries with a $\delta$~Sct component are powerful tools to derive the fundamental parameters 
and probe the internal structure of stars. In this study, spectral analysis of 6 primary
$\delta$~Sct  components in eclipsing binaries has been performed. Values of \teff,
\vsini, and metallicity for the stars have been derived from medium-resolution spectroscopy. Additionally, a revised list of $\delta$~Sct 
stars in eclipsing binaries is presented. In this list, we have only given the $\delta$~Sct
stars in eclipsing binaries to show the effects of the
secondary components and tidal-locking on
the pulsations of primary $\delta$~Sct components. The stellar pulsation, atmospheric and
fundamental parameters (e.g., mass, radius) of 92 $\delta$~Sct stars in eclipsing
binaries have been gathered. Comparison of the properties of single and eclipsing binary member $\delta$
Sct stars has been made. We find that single $\delta$~Sct stars pulsate in longer periods and with higher
amplitudes than the primary $\delta$~Sct components in eclipsing binaries. The {\vsini} of $\delta$~Sct components is
found to be significantly lower than that of single $\delta$~Sct stars. Relationships between
the pulsation periods, amplitudes, and stellar  parameters in our list have been examined. 
Significant correlations between the pulsation periods and the orbital periods, \teff, \logg,
radius, mass ratio, \vsini, and the filling factor have been found.
\end{abstract}

\begin{keywords}
Stars: binaries: eclipsing -- stars: fundamental parameters -- stars: variables: $\delta$ Scuti
\end{keywords}

\section{Introduction}

The $\delta$~Scuti ($\delta$~Sct) stars are remarkable objects  for asteroseismology particularly because
of their pulsation mode variability. The $\delta$~Sct stars oscillate in low-order radial and
non-radial pressure and gravity modes and most of them have frequency range of $5-50$ d$^{-1}$
\citep{2000ASPC..210....3B}. Pulsations are driven by the $\kappa$-mechanism in these  variables
\citep{1999A&A...351..582H}.  The $\delta$~Sct stars are dwarf to giant stars with spectral types between 
A0 and F0 \citep{2013AJ....145..132C}. These variables have masses from 1.5 to 2.5\,$M_{\rm \odot}$
and are located on or near the main sequence \citep{2010aste.book.....A}. Therefore, they are in a transition region where the
convective envelope turns to a radiative envelope, while energy starts to be transferred by
convection in the core of the star \citep{2010aste.book.....A}. The $\delta$~Sct
stars allow us to understand the processes occurring in this transition region by using their pulsation
modes.

Approximately 70\% of stars are binary or multiple systems
\citep{2009AJ....137.3358M,2011IAUS..272..474S,2014MNRAS.443.3022A}. Therefore, it is likely to find a
$\delta$~Sct variable as a member of a binary system. The existence of a pulsating variable in an eclipsing
binary system makes this variable more valuable. Using the pulsation characteristic, the interior structure of
the star can be probed and, using the eclipsing characteristic, the fundamental parameters (e.g.\, mass, radius) of
the pulsating component can be derived by modelling the light and radial velocity curves of a binary system. 
These fundamental parameters are important to make a reliable model of a pulsating star. Thus, the
interior structures and the evolution statuses of stars can be examined in detail.

Many $\delta$~Sct stars in eclipsing binary systems have been discovered
\citep[e.g.][]{2016MNRAS.460.4220L,2016NewA...46...40S}. A group of eclipsing binaries with a $\delta$~Sct
component was defined as oscillating eclipsing Algol (oEA) systems by \citet{2004A&A...419.1015M}. The oEA
systems are B to F type mass-accreting main-sequence pulsating stars in semi-detached 
eclipsing binaries. Because of mass-transfer from the secondary components onto the primary pulsating
stars and also due to the tidal distortions in oEA systems, the pulsation parameters and the evolution of primary
pulsating components can be different.

There have been several studies on the effect of binarity on $\delta$~Sct type pulsations. Firstly,
\citet{2006MNRAS.366.1289S} showed the effect of orbital period on the pulsation period.  The relation between 
orbital and pulsation periods was
theoretically revealed by \citet{2013ApJ...777...77Z}. They showed that pulsation periods vary 
depending on the orbital period, mass ratio of binary system and filling factor of the primary
pulsating component. It was also shown that the gravitational force applied by 
secondary components onto their primary components influences the pulsation periods of primary $\delta$~Sct
components \citep{2006MNRAS.366.1289S}. Because of the effects of mass-transfer and tidal
distortions in semi-detached binaries, the primary $\delta$~Sct components also evolve more slowly
through the main sequence than single $\delta$~Sct stars \citep{2015ASPC..496..195L}. 

The number of known binaries with a $\delta$~Sct component constantly increases. 
Additionally, hybrid stars, which show both $\delta$~Sct and $\gamma$~Dor type pulsations, 
have been discovered in eclipsing binary systems \citep{2015A&A...584A..35S, 2013MNRAS.434..925H}.
In a recent study, an updated list of $\delta$~Sct stars in binaries was presented by \citet{2017MNRAS.465.1181L}. 
In their study, all known $\delta$~Sct stars and also $\delta$~Sct\,-\,$\gamma$~Dor hybrids 
in binaries were collected, including the non-eclipsing
ones. Although 199 binary systems are given in their list, there are only 87 detached and semi-detached eclipsing
binaries containing a $\delta$~Sct variable. 
The others are mostly visual binaries, ellipsoidal variables, and spectroscopic binaries in which the
fundamental parameters cannot be derived as precisely as in eclipsing binary systems.

As listed by \citet{2006ASPC..349..153L}, some open questions about the eclipsing binaries with a
pulsation component exist. The effect of binarity on pulsation quantities (period and amplitude), possible
connections between orbital motion, rotation, chemical composition, and pulsation are some of these
questions. Therefore, we have focused on the eclipsing binary systems with $\delta$~Sct components in this
study. To obtain the stellar atmospheric parameters, a spectroscopic analysis of six $\delta$~Sct stars in
eclipsing binary systems has been performed. A revised list of $\delta$~Sct stars in eclipsing binaries is 
presented to show the effects of secondary components and fundamental stellar parameters on the pulsations of
the primary pulsating components.

Information about the spectroscopic observations and data reduction are given in Sect\,2. The 
spectroscopic analysis of the stars is presented in Sect\,3. The revised list of eclipsing binary systems 
with a $\delta$~Sct component, general properties of these systems, and the relations between
pulsation  periods, amplitudes and fundamental parameters of the stars are introduced in
Sect\,4. In Sect\,5 we present a discussion on the correlations found, the positions of
$\delta$~Sct stars in eclipsing binaries in the $\log {\teff}$\,--\,{\logg} diagram, and a comparison of the
properties of single and eclipsing binary members $\delta$~Sct stars. The conclusions are given in Sect\,6.

\begin{table}
\centering
  \caption{Information about the spectroscopic survey. S/N gives the values for combined spectra 
  apart from CL Lyn. The S/N of CL Lyn is the value for ELODIE spectrum.}
    \label{ondspe}
  \small
  \begin{tabular}{lcccc}
  \hline
  Name  & Observation     & S/N  & Number of & Light \\
        & dates            &      & spectra   & Contribution$^{a}$\\
 \hline
 XX Cep   & 2015-06/09    & 50 & 2 & 3.7\,$^{[1]}$\\
 UW Cyg   & 2015-07/09    & 35 & 2 & 5.1\,$^[2]$\\    
 HL Dra   & 2015-06/07    & 120& 2 & 4.4\,$^[2]$ \\
 HZ Dra   & 2015-06/07/09 & 80 & 3 & 0.5\,$^[2]$ \\
 TZ Dra   & 2015-06/09	  & 60 & 2 & 7.6\,$^[3]$ \\
 CL Lyn   & 2015-09\&2001 & 50 & 2 & 4.2\,$^[2]$\\
\hline
\end{tabular}
\begin{description}
 \item[$^{a}$] Percentage of light contribution of secondary component in $B$ band.
 [1] \citet{2016AJ....151...77K}, [2] \citet{2012MNRAS.422.1250L}, [3] \citet{2013Ap&SS.343..123L}
\end{description}
\end{table}

\section[]{Observations}

Spectroscopic observations of six eclipsing binaries with a primary $\delta$~Sct component were
carried out. The stars were selected taking into account the secondary components' light contributions,
in order to obtain spectra which are less influenced by the light of secondary
components. The light contributions of the stars from literature photometric analyses are given in Table\,\ref{ondspe}.

The observations were carried out using the 2-metre Perek Telescope at the Ond\v{r}ejov Observatory (Czech
Republic). We acquired spectra with the coud\'e slit spectrograph at its 700-mm focus, in which the PyLoN
2048$\times$512 BX CCD  chip was used \citep[for details, see][]{2002PAICz..90....9S}. 
The resolving power of the instrument is about 25\,000 at 4300\,{\AA}. The spectra were
taken in the wavelength range of 4272--4506\,{\AA}, which covers the H$\gamma$ line. This wavelength region was also
selected because metal lines (e.g. \ion{Ti}, \ion{Mg}\,and \ion{Fe}) are more numerous in this range of effective temperature.

To further minimise the light contribution of secondary components in the spectra, spectra of each
star were taken at approximately 0.5 orbital phase when the primary is covering the secondary. 
The individual spectra were combined to increase the
signal-to-noise (S/N) ratio. For CL Lyn an
ELODIE\footnote{\url{http://atlas.obs-hp.fr/elodie/}} spectrum, taken
in 2001, was used in addition to our observation. Information about the spectroscopic survey is given in
Table\,\ref{ondspe}. The stars are semi-detached eclipsing binaries with a primary
$\delta$~Sct component, except for HZ Dra which is a detached binary with a primary
$\delta$~Sct component \citep{2012MNRAS.422.1250L}.

The reduction and normalisation of the spectra were performed using the NOAO/IRAF
package\footnote{\url{http://iraf.noao.edu/}}. In the reduction process, bias subtraction, flat-field
correction, scattered light extraction and wavelength calibration were applied. The reduced ELODIE
spectrum for CL Lyn was used. The standard reduction was performed by the dedicated reduction pipeline of
ELODIE. The spectra of each star were manually normalised using the \textit{continuum} task of the
NOAO/IRAF package.

\section{Spectroscopic analysis}

Prior to detailed spectroscopic analysis, spectral classifications of the stars were
obtained. The effective temperature (\teff) of the primary components were derived using the 
spectral energy distribution (SED) and the
H$\gamma$ line. The metallicities were obtained using the spectrum fitting method.

\begin{table*}
\caption{The stellar parameters of the six $\delta$~Sct stars in eclipsing binaries.}            
\label{ondparam}     
\centering     
\footnotesize
\begin{tabular}{llcllllllc}
\hline                
  Name   & V      & {$E(B-V)$}      &    Sp type    &    Sp type  & {\teff}          &  {\teff}         & {\logg} $^{a}$ &  $v\sin i$ & $[m/H]$\\
         &(mag)   & (mag)           & (literature)  &(This Study) &  (SED)                   &  (Spec)          &              & (\kms)    &    \\
         &        &  $\pm$\,0.023   &               &             & (K)              &  (K)             &                &           & \\
         &        &                 &               &             &                  &                  &                &           &       \\
   \hline
 HL Dra  & 7.36   & 0.040           & A5$^{[1]}$    & A6\,IV      & 7786\,$\pm$\,174 & 7800\,$\pm$\,200 & 4.22           & 107\,$\pm$\,10   & $-$0.12\,$\pm$\,0.17\\
 HZ Dra  & 8.14   & 0.016           & A0$^{[1]}$    & A8/A7\,V    & 7926\,$\pm$\,250 & 7700\,$\pm$\,200 & 4.07           & 120\,$\pm$\,10   & $-$0.09\,$\pm$\,0.20\\
 XX Cep  & 9.18   & 0.026           & A6\,V$^{[2]}$ & A7\,V       & 7160\,$\pm$\,152 & 8200\,$\pm$\,300 & 4.09           & 54\,$\pm$\,5     & \,\,\,\,0.59\,$\pm$\,0.23\\
 TZ Dra  & 9.32   & 0.020           & A7\,V$^{[3]}$ & A7\,V       & 7382\,$\pm$\,173 & 7800\,$\pm$\,200 & 4.26           & 86\,$\pm$\,8     & $-$0.01\,$\pm$\,0.22\\
 CL Lyn  & 9.77   & 0.181           & A5$^{[1]}$    & A8\,IV      & 7699\,$\pm$\,189 & 7600\,$\pm$\,300 & 3.98           & 75\,$\pm$\,3     & $-$0.16\,$\pm$\,0.20\\
 UW Cyg  & 10.86  & 0.101           & A6V$^{[4]}$   & A7/A6\,IV   & 7550\,$\pm$\,176 & 7800\,$\pm$\,350 & 4.06           & 45\,$\pm$\,10    & \,\,\,\,\,$^{*}$\\
\hline                        
\end{tabular}

\begin{description}
 \item[]$^{a}$Calculated using the stars' masses and radii which are given in Table\,\ref{delsctecl}. 
 $^{*}$ Could not be calculated because of the low S/N ratio. [1] \citet{1997yCat.1239....0E}, [2] \citet{2016AJ....151...77K}, [3] \citet{1960ApJ...131..632H}, [4] \citet{2012MNRAS.422.1250L}
\end{description}
\end{table*}

\subsection{Spectral classification}

Preliminary information about the atmospheric parameters (\teff, surface gravity \logg) 
and surface peculiarities of stars can be obtained by spectral classification \citep{2014dap..book.....N}.

The spectral and luminosity types of stars are identified by comparing their spectra with a group of 
well-known standard stars' spectra. The A--F type standard stars were used in our classification
\citep{2003AJ....126.2048G},  because the $\delta$~Sct variables are A and F type stars. The spectral
types of the stars were derived primarily using H$\gamma$ and neutral metal lines
\ion(Fe, Ti) in the 4400--4500\,{\AA} wavelength region. The luminosity types were also derived
by using the ionised metal lines. 

The spectral and luminosity types obtained for the stars are given in Table\,\ref{ondparam}. Only for HZ Dra, newly
determined spectral classification (A8/A7\,V) was found to be significantly different  than the
previous classification (A0, \citealt{1975ascp.book.....H}), 
while the other spectral types are mostly in agreement with the literature.

\subsection{Determination of \teff, \vsini, and metallicity}

The {\teff} of the primary $\delta$~Sct components were obtained from the H$\gamma$ line 
and metallicities were determined using metal lines in the 4400--4500\,{\AA} wavelength region.

Prior to the spectral analysis, we determined {\teff} from the SED. The SEDs were constructed from literature photometry and spectrophotometry,
using 2MASS $J$, $H$ and $K_s$ magnitudes \citep{2006AJ....131.1163S}, Tycho $B$ and $V$ magnitudes
\citep{1997A&A...323L..57H}, USNO-B1 $R$ magnitudes \citep{2003AJ....125..984M}, TASS $I$ magnitudes \citep{2006PASP..118.1666D}, 
and data from the Ultraviolet Sky Survey Telescope
(TD1) \citep{1973MNRAS.163..291B}. However, TD1 data are only available for HL Dra and HZ Dra.
To remove the effect of interstellar reddening on the SED, $E(B-V)$ values 
were calculated from the Galactic extinction maps \citep{2005AJ....130..659A}, with distances obtained from
Gaia parallaxes \footnote{\url{https://gea.esac.esa.int/archive/}} \citep{2016arXiv160905175C}. 
The ${\it E(B-V)}$ values are given in Table\,\ref{ondparam}. The average uncertainty in $E(B-V)$ 
was found to be 0.023\,mag. The SEDs were de-reddened using the analytical
extinction fits of \cite{1979MNRAS.187P..73S} for the ultraviolet and
\cite{1983MNRAS.203..301H} for the optical and infrared.

\begin{figure}
\includegraphics[width=8cm,angle=0]{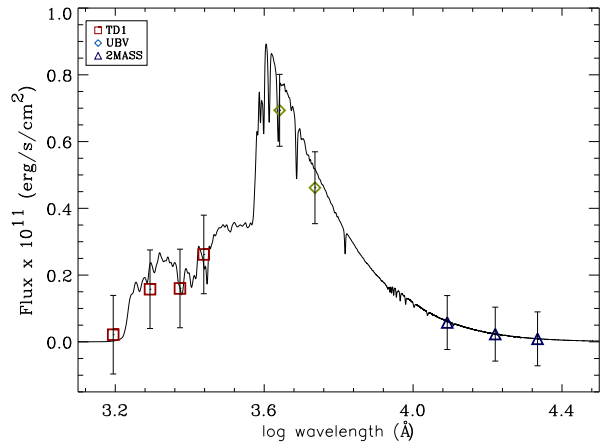}
\caption{SED fit for HL Dra.}
\label{figure_ek}
\end{figure}

The stellar \teff\,(SED) values were determined by fitting solar-composition
\cite{1993KurCD..13.....K} model fluxes to the de-reddening SEDs. The model
fluxes were convolved with photometric filter response functions. A weighted
Levenberg-Marquardt non-linear least-squares fitting procedure was used to find
the solution that minimised the difference between the observed and model
fluxes. Since {\logg} is poorly constrained by our SEDs, we fixed {\logg}~=~4.0 
for all the fits. The uncertainties in \teff\,(SED) includes the formal least-squares error and
uncertainties in $E(B-V)$ (0.023) and $\log g$ (0.2) added in quadrature. In Fig.\,\ref{figure_ek}, we show the SED fit for HL Dra.

By using the initial values of \teff\,(SED), fundamental atmospheric parameters were determined for each star. 
Before the hydrogen line analysis and metallicity determinations of the stars, the projected rotational velocity (\vsini) 
values were derived. A theoretical spectrum of each star was calculated 
using the initial atmospheric parameters. The metal lines in the spectrum of each star
were matched to the theoretical spectrum by adjusting {\vsini} \citep{2008oasp.book.....G}. 
Final values of {\vsini} were obtained 
by minimising the difference between the observed and 
theoretical spectra. The {\vsini} values 
are given in Table\,\ref{ondparam}.

\begin{figure}
\includegraphics[width=8cm,angle=0]{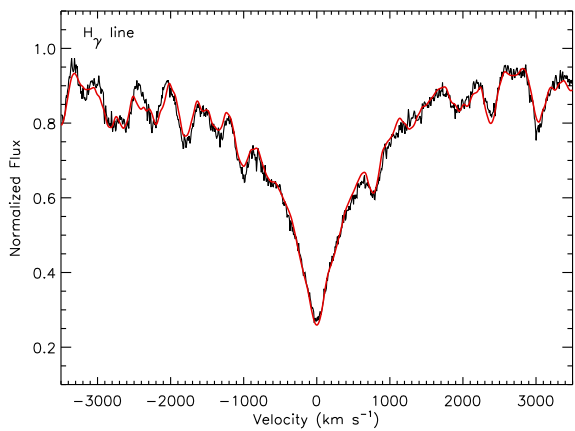}
\caption{Comparison of the calculated (red line) and observed spectra of H$\gamma$ line of HL Dra.}
\label{figure1}
\end{figure}

Hydrogen lines are good temperature indicators, since they are insensitive to {\logg} for stars with {\teff}~$<$~8000\,K
\citep{2008oasp.book.....G, 2002A&A...392..619H}. In our analysis, we adopted value of {\logg} which were calculated using
the stars' masses and radii values given in the literature (see Table\,A1). Additionally, {\vsini} and metallicity ($[m/H]$, assumed to be solar) 
were fixed. In our analysis, the hydrostatic, plane-parallel, 
local thermodynamic equilibrium ATLAS9 models \citep{1993KurCD..13.....K} were used and the SYNTHE code \citep{1981SAOSR.391.....K} was used 
to produce theoretical spectra. {\teff} was determined by minimising the difference between the theoretical and observed
hydrogen line, as described by \citet{2004A&A...425..641C}. 
The {\teff}\,(Spec) values obtained are listed in Table\,\ref{ondparam} and
comparison of the calculated and observed spectra for one of the stars is shown in  Fig.\,\ref{figure1}. 

\begin{figure}
\includegraphics[width=8.4cm, angle=0]{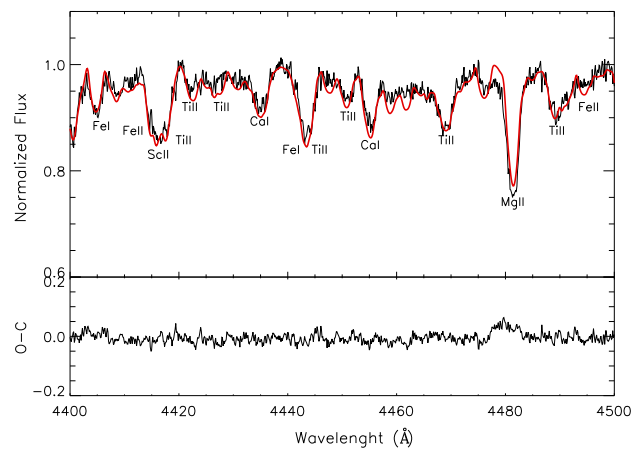}
\caption{Upper panel: comparison of the theoretical (red lines) and observed spectra of HL Dra. 
Lower panel: difference between the observed and theoretical spectra.}
\label{figure2}
\end{figure}

The $[m/H]$ values were derived using the {\teff}\,(Spec), {\logg} and {\vsini} values given in Table\,2. The analysis was executed using 
version 412 of the Spectroscopy Made Easy (SME) package \citep{1996A&AS..118..595V}, which determines atmospheric parameters, 
$[m/H]$ and elemental abundances using the spectrum fitting method. 
In this analysis, atmosphere models produced by ATLAS9 code \citep{1993KurCD..13.....K} were used. 
The line list 
was taken from the Vienna Atomic Line
Database\footnote{\url{http://vald.astro.uu.se/}} \citep[VALD][]{1995A&AS..112..525P}.
The $4400-4500$\,{\AA} wavelength range was used in the metallicity analysis. The $[m/H]$ values obtained are given in Table\,\ref{ondparam}. 
However, $[m/H]$ could not be determined for UW Cyg because of the low S/N ratio of the spectra. The comparison of the 
theoretical and observed spectra used in the $[m/H]$ analysis is demonstrated in Fig.\,\ref{figure2}.
Uncertainties of the spectroscopic parameters comprise the least-squares error 
and the uncertainties caused by the fixed parameters in each analysis. Additionally, the error due to low 
S/N ratio was included, using the value from \citet{2016MNRAS.458.2307K}.

The {\teff} values of the analysed stars were also compared with those used in the literature. 
In this comparison, we noticed that if {\teff} had been obtained from spectral
classifications, the {\teff} used in previous studies are in agreement with our spectroscopic results 
to within the errors. Others, however, have completely different values. We determined {\teff}
values of HZ Dra and XX Cep to be 7700 and 8200\,K, respectively. However, their previously used \teff
values are 9800 K for HZ Dra \citep{2012MNRAS.422.1250L}
and 7300~K for XX Cep  \citep{2014NewA...27...95H}. However, a newer spectral analysis
of XX Cep was recently presented by \citet{2016AJ....151...77K}. They obtained {\teff} =
7946\,$\pm$\,240\,K and {\vsini} = 48.6\,$\pm$\,6.8\,{\kms}, which are in good agreement with our
results.

\section{$\delta$~Sct stars in Eclipsing Binary systems}

The properties of $\delta$~Sct stars in eclipsing binaries can be
different from single $\delta$~Sct stars. Especially, the pulsating primary components
in close binary systems evolve differently than single ones \citep{2015ASPC..496..195L}. In close binary
systems, the primary component can gain mass from the secondary and can be covered
by material from the secondary. Additionally, tidal distortion is present in
these systems. Mass-transfer and tidal distortion will affect the pulsation
period ($P_{\rm puls}$) and amplitude ($Amp$) of $\delta$~Sct stars in close
binaries. How much the binarity affects the $\delta$~Sct pulsations
in binaries is one of the open questions.

To show the effect of binarity on pulsation, the correlations between $P_{\rm puls}$, $Amp$ and
the orbital and atmospheric parameters of eclipsing binary member $\delta$~Sct stars have been
examined. Firstly, a correlation between $P_{\rm puls}$ and orbital period ($P_{\rm orb}$) for 20
$\delta$~Sct stars in eclipsing binaries was found by \citet{2006MNRAS.366.1289S}. The
$P_{\rm puls}$\,--\,$P_{\rm orb}$ correlation was improved by newer discoveries
\citep{2012MNRAS.422.1250L}. Then, a theoretical explanation for the $P_{\rm puls}$\,--\,$P_{\rm orb}$
correlation was  given by \citet{2013ApJ...777...77Z} who expressed $P_{\rm puls}$ mainly as a
function of the pulsation constant (Q), the filling factor ($f$), 
$P_{\rm orb}$, and the mass ratio ($q=M_\text{secondary}/M_\text{primary}$, where $M$ denotes the mass)
with the following equation:

\begin{equation} \label{eq1}
P_\mathrm{puls}  = \frac{G^{1/2}}{2\pi}Qf^{3/2}r^{3/2}(1+q)^{1/2}P_\mathrm{orb}
\end{equation}

\noindent where $G$ and $r$ are the gravitational constant and effective radius (radius divided by
semi-major axis), respectively. $f$ shows how much a star fills its Roche potential ($\Omega$) and it is expressed by: 

\begin{equation} \label{eq2}
f= (\Omega_{\rm inner} - \Omega) / (\Omega_{\rm outer} - \Omega_{\rm inner}),
\end{equation}

\citet{2013ApJ...777...77Z} tested whether this 
theoretical approach is compatible with the observed correlation of $P_{\rm puls}$\,--\,$P_{\rm orb}$ using 69 
eclipsing binaries with $\delta$~Sct stars. They found that the theoretical correlation is in agreement with the observed one. 
This correlation was also confirmed by \citet{2017MNRAS.465.1181L}. They obtained a similar 
correlation using 66 semi-detached and 25 detached systems which have $P_{\rm orb}$ $\leqslant$ 13 days. 
They also showed that for binaries with $P_{\rm orb}$ $>$ 13 days, there is no significant effect of binarity 
on pulsations. In their study, the known correlation between $P_{\rm puls}$ and {\logg} was also shown for 82 systems 
which contained semi-detached, detached and unclassified stars with $P_{\rm orb}$ $<$ 13 days. 
However, it should be kept in mind that in their study, both eclipsing and non-eclipsing 
binaries were used and some of these stars were assumed to be detached systems. Additionally, 
a negative correlation between $P_{\rm puls}$ of primary $\delta$~Sct components and gravitational force applied by secondary components 
onto the pulsating stars has been found \citep{2006MNRAS.366.1289S, 2012MNRAS.422.1250L, 2017MNRAS.465.1181L}. 

The known and possible correlations between the fundamental absolute parameters (e.g., masses,
radii), atmospheric parameters, and  the pulsation quantities ($P_{\rm puls}$, $Amp$) of $\delta$~Sct
components in eclipsing binary systems give us an opportunity to understand these stars in detail and
show the effect of secondary components, mass transfer and tidal locking on pulsation quantities. 
Therefore, we have prepared a revised list of eclipsing binaries with a primary $\delta$~Sct
component.  The list includes 67 semi-detached and 25 detached eclipsing binaries. Seven of these
stars (WX Dra,  GQ Dra, KIC\,06669809, KIC\,10619109, KIC\,1175495, KIC\,10686876 and KIC\,6629588)
were taken from \citet{2016arXiv160608638L,2017MNRAS.465.1181L}. In these studies, the stars were
found to be $\delta$~Sct variables in eclipsing binaries for the first time. In our study, we did not
include any stars which have unclassified Roche geometry.

The parameters of the primary and secondary components of our sample of 92 eclipsing binary systems with a
primary $\delta$~Sct component were gathered from the literature. In this revised list, values of
\teff, \logg, masses ($M$), radii ($R$),  luminosities ($L$), bolometric magnitudes (M$_{\rm bol}$),
semi-major axis ($a$) of the primary and secondary components and $f$, \vsini,
$P_{\rm puls}$, peak-to-peak $Amp$ in $V$ and $B$-band of primary pulsating components and the parallaxes, orbital inclinations ($i$), and 
the $q$ of binary systems were collected, as well as the basic parameters of the systems
such as visual magnitudes ($V$), spectral types (SP) and binary types. This updated list
contains more stars and a wider variety of stellar parameters compared to the previous list
\citep{2017MNRAS.465.1181L}. The updated list is given in Table\,\ref{delsctecl}.

\subsection{General properties of $\delta$~Sct components in eclipsing binaries}

In our list, there are only two eclipsing binaries that show $\delta$~Sct
type pulsations in both components (RS Cha and KIC\,09851944). In the
other eclipsing binaries, only the primary components exhibit $\delta$~Sct type
pulsations. Therefore, in all our examinations we only took into account the
properties of the primary $\delta$~Sct components.

\begin{figure}
\includegraphics[width=7.3cm,height=8cm, angle=0]{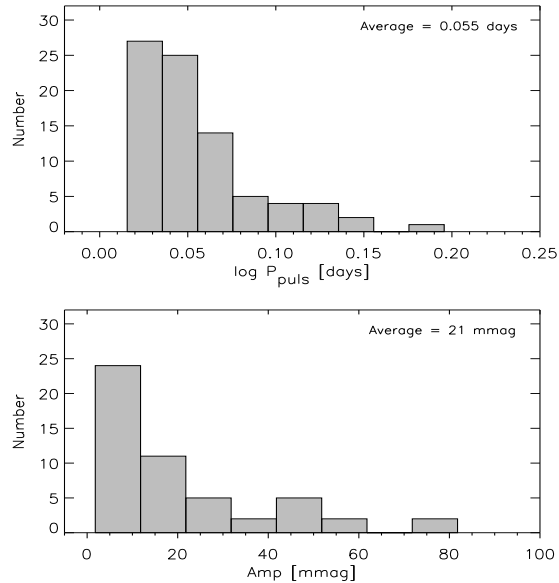}
\caption{The distributions of $P_{\rm puls}$ and $V$-band $Amp$ of the primary $\delta$~Sct components in eclipsing binary systems.}
\label{figure3}
\end{figure}

The $P_{\rm puls}$ and $V$-band $Amp$ distributions of primary $\delta$~Sct
components in eclipsing binaries are shown in  Fig.\,\ref{figure3}. The pulsation
periods of the highest amplitudes were collected in the list and only $V$-band $Amp$ of the stars 
was used in any comparisons and analyses in this study. It is clearly seen that $\delta$~Sct
type primary pulsating components mainly oscillate in periods between
$\sim$0.016 and 0.195 days, with an average amplitude of 21\,mmag. In
the list, there is also a high-amplitude $\delta$~Sct star (HADS) (V1264 Cen)
which is not used in  Fig.\,\ref{figure3} and excluded in the next steps. The
average $P_{\rm puls}$ values for semi-detached and detached systems were
found to be 0.049 and 0.073 days, respectively. Although
the number of detached systems is lower than the semi-detached ones,
there is a clear distinction between the pulsation periods of both systems.
However, we did not find a significant difference in the $V$-band $Amp$ of
$\delta$~Sct components of both type of eclipsing binary. The reason for
lower pulsation periods of primary $\delta$~Sct components in the semi-detached
systems can be the effects of tidal locking and mass-transfer from the
secondary non-pulsating component to the primary pulsating component \citep{2012MNRAS.422.1250L, 2006MNRAS.370.2013S, 2003AJ....126.1933S}. 
The primary component gains mass from the secondary component and this can change
the surface composition and internal structure of the primary pulsating component
and also the angular momentum of both components changes during this process \citep{2010aste.book.....A}.
Hence, mass-transfer could affect the oscillations. 

\begin{figure*}
\includegraphics[width=16cm,height=6.5cm, angle=0]{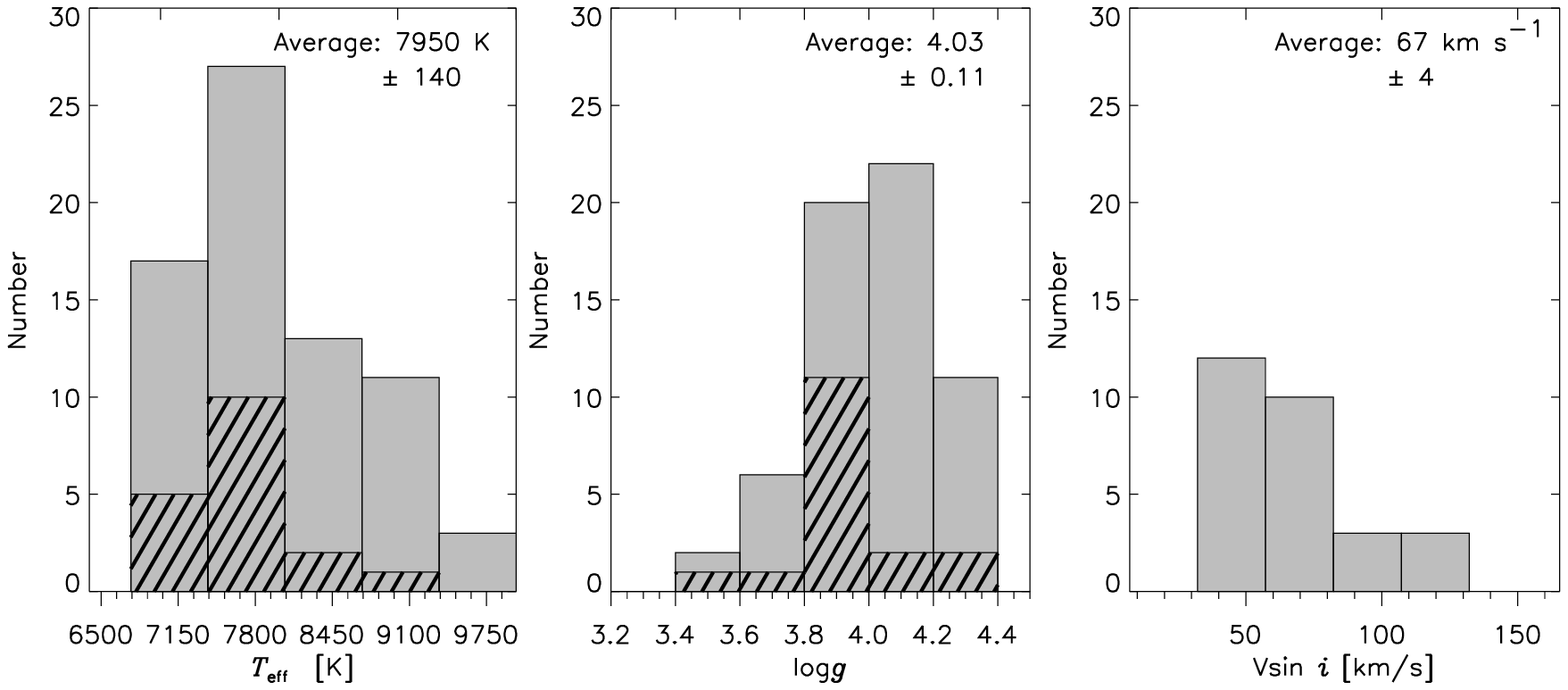}
\caption{The distributions of \teff, \logg, and {\vsini} values of primary $\delta$~Sct components in eclipsing binaries. The gray histograms 
show the distributions of whole sample, while slanted lines represent the distributions of the stars that have {\teff} and {\logg} values derived by the spectroscopic 
analyses}
\label{figure4}
\end{figure*}

\begin{figure}
\includegraphics[width=8cm, angle=0]{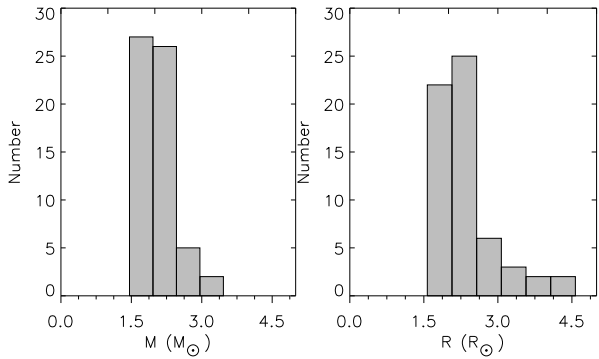}
\caption{The distributions of $M$ and $R$ values of primary $\delta$~Sct components in eclipsing binaries.}
\label{figure5}
\end{figure}

The \teff, {\logg}, and {\vsini} ranges of $\delta$~Sct components in eclipsing
binary systems are  illustrated in  Fig.\,\ref{figure4}. In the figure, the distributions of
parameters obtained from photometric  and spectroscopic studies are shown. Both
photometric and spectroscopic {\teff} and {\logg} values have similar ranges,
which are 6750\,--\,9660\,K and 3.40\,--\,4.38, respectively. The {\logg}
values of semi-detached and detached systems are in the range of 3.80\,--\,4.38
and 3.50\,--\,4.20, respectively, while {\teff} ranges of these eclipsing
binary types are similar. One semi-detached system, QY Aql, has a
{\logg} value of 3.40 \citep{2013Ap&SS.343..123L}, which is low
for the primary component of a semi-detached system, since they are generally
main-sequence stars. The {\vsini} values of the primary $\delta$~Sct
components were found to be in the range from 12 to 130\,{\kms}. The values of $M$ and $R$
for the primary $\delta$~Sct components were also found in the ranges of
$1.46-3.30$\,$M_{\sun}$ and $1.57-$4.24\,$R_{\sun}$, respectively, as shown in  Fig.\,\ref{figure5}. 
No significant difference was obtained between the range of $M$ and $R$
values for detached and semi-detached systems.

\subsection{Correlations between the collected parameters and the pulsation quantities}

\begin{figure}
\includegraphics[width=8.5cm, angle=0]{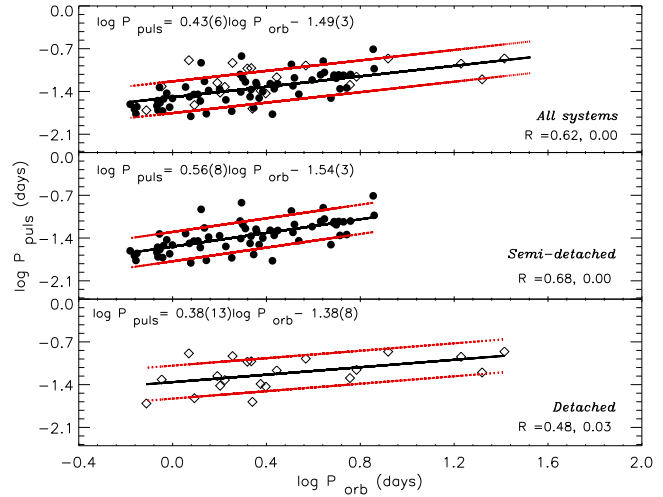}
\caption{The correlations between $P_{\rm puls}$ and $P_{\rm orb}$ for detached (lower panel), semi-detached (middle panel) and 
all systems (upper panel). The filled circles, diamonds and red lines 
represent the semi-detached, detached and 1-$\sigma$ levels, respectively. The equations in each panel were obtained from the correlations.
R constant shows the spearman rank which gives the strength of correlation (number before the comma in the R constant) and the deviation amount of points from the 
correlation (number after the comma in the R constant).}
\label{figure6}
\end{figure}

The known and possible correlations between the collected 
fundamental, atmospheric and orbital parameters, and the
pulsation quantities of the primary $\delta$~Sct components in eclipsing binary
systems were examined. Firstly, the known correlation between $P_{\rm puls}$ and
$P_{\rm orb}$ was checked for semi-detached and detached systems. These 
correlations are demonstrated in Fig.\,\ref{figure6}. Average errors in
$P_{\rm puls}$ and $P_{\rm orb}$ are  about 10$^{-3}$ and 10$^{-5}$ days, and the error bars
are smaller than the size of the symbols. Therefore, the error bars of
$P_{\rm puls}$ and $P_{\rm orb}$ are not shown in this and subsequent figures.
Additionally, for some stars, the errors of the parameters were not given in the
literature. Therefore, the average uncertainties of the parameters are shown
in all figures. 

\begin{figure}
\includegraphics[width=8.5cm, angle=0]{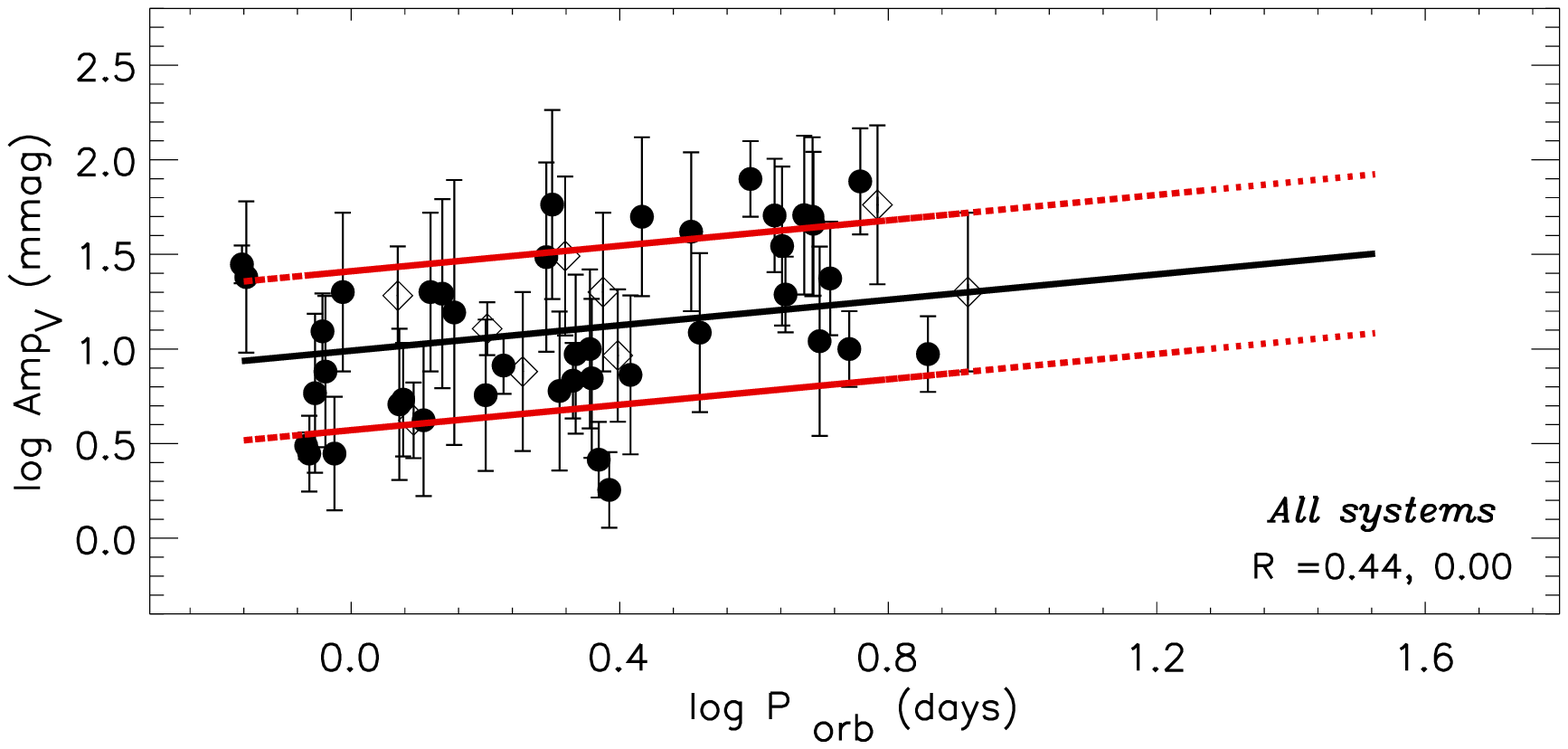}
\caption{The correlation between the V-band pulsation $Amp$ and $P_{\rm orb}$ of primary $\delta$~Sct stars in eclipsing binary systems. 
The symbols, lines, and the R constant are the same as in Fig.\,\ref{figure6}.}
\label{figure7}
\end{figure}

Significant positive $\log P_{\rm puls}$\,--\,$\log P_{\rm orb}$ correlations were found
for both semi-detached and detached systems. The relationships for these
correlations are given in the top of each panel in  Fig.\,\ref{figure6}. The correlation for
semi-detached systems was found to be stronger than for the detached systems. As
can be seen from  Fig.\,\ref{figure6}, all stars are mainly inside the 1-$\sigma$ level.
The $V$-band pulsation $Amp$ relation with $P_{\rm orb}$ was examined as well.  As a
result, a correlation was found between these parameters as shown in  Fig.\,\ref{figure7}.
However, the correlation is not strong, because of the scatter and number of data of points.

\begin{figure}
\includegraphics[width=8.5cm,height=8.7cm,  angle=0]{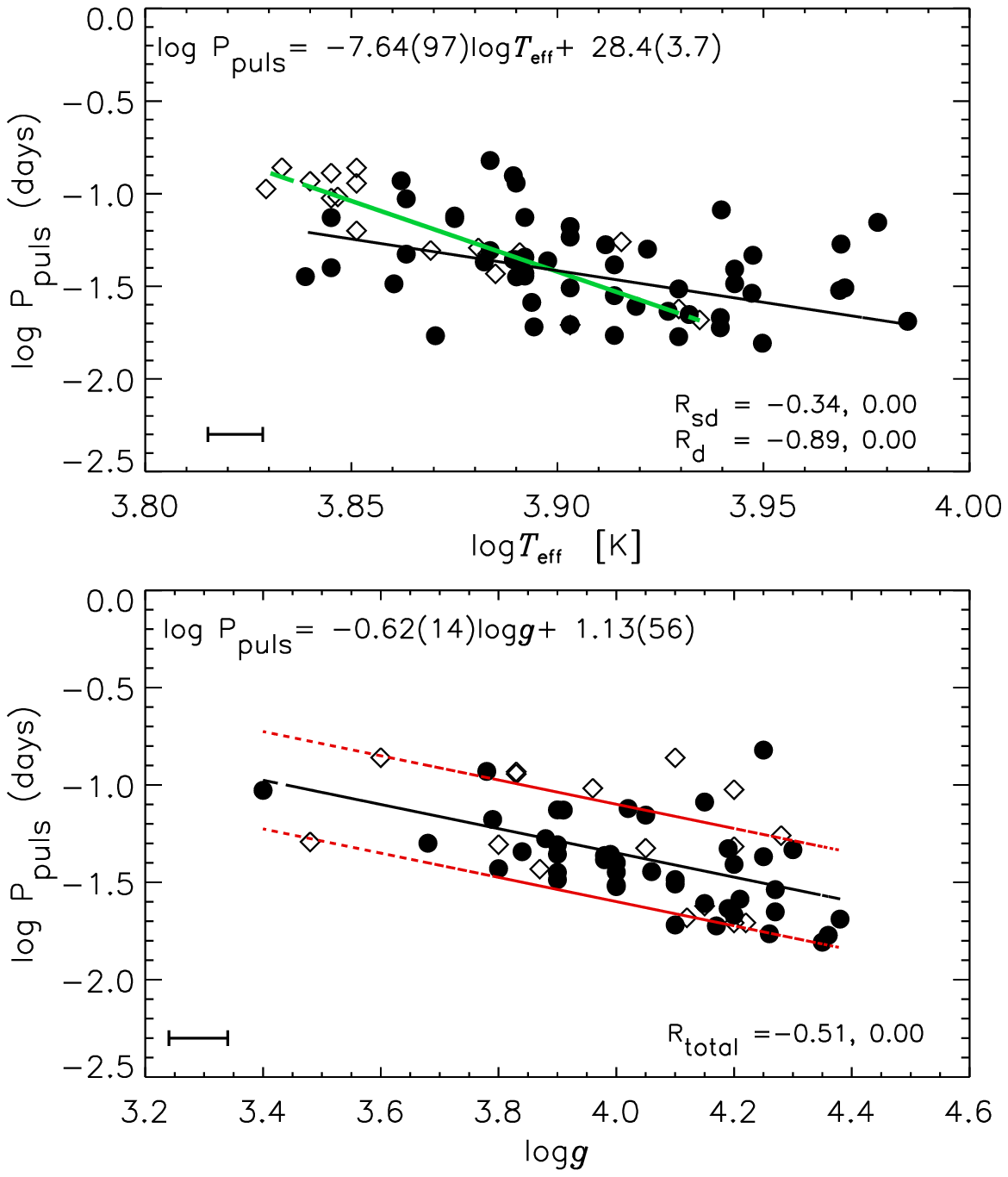}
\caption{The $\log {\teff}$\,--\,$\log P_{\rm puls}$ (upper panel) and {\logg}\,--\,$\log P_{\rm puls}$ (lower panel) correlations. Green 
line in upper panel illustrates the correlation only for detached systems, while black line shows the correlation of semi-detached systems. 
The equations in upper and lower panels 
are given for detached and all systems' primary $\delta$~Sct components considering the correlations, respectively. 
The symbols, red lines, and R constant are the same as in Fig.\,\ref{figure6}.}
\label{figure8}
\end{figure}

The correlations between the atmospheric parameters (\teff, \logg) and
$P_{\rm puls}$ and $V$-band $Amp$ of primary $\delta$~Sct components were
examined. While no correlation between $Amp$ and the
atmospheric parameters was found, there are significant correlations between
\teff, \logg, and $P_{\rm puls}$. As shown in  Fig.\,\ref{figure8}, these correlations were found
for all types  of eclipsing binaries' primary $\delta$~Sct components
and they show a negative variation in $P_{\rm puls}$ with increasing {\teff} and \logg.
However, as can be seen from the upper panel of  Fig.\,\ref{figure8}, the
log\,{\teff}\,--\,$\log P_{\rm puls}$ relation is stronger for the
pulsating primary components of detached systems than for semi-detached
systems. Therefore, only the relationship for the correlation for
detached systems is given in  Fig.\,\ref{figure8}. The {\logg}\,--\,$\log P_{\rm puls}$
relationship and the correlation for the primary $\delta$~Sct
components in all types of eclipsing binary systems are also shown in Fig.\,\ref{figure8}. 

\begin{figure}
\includegraphics[width=8.5cm,height=8.7cm, angle=0]{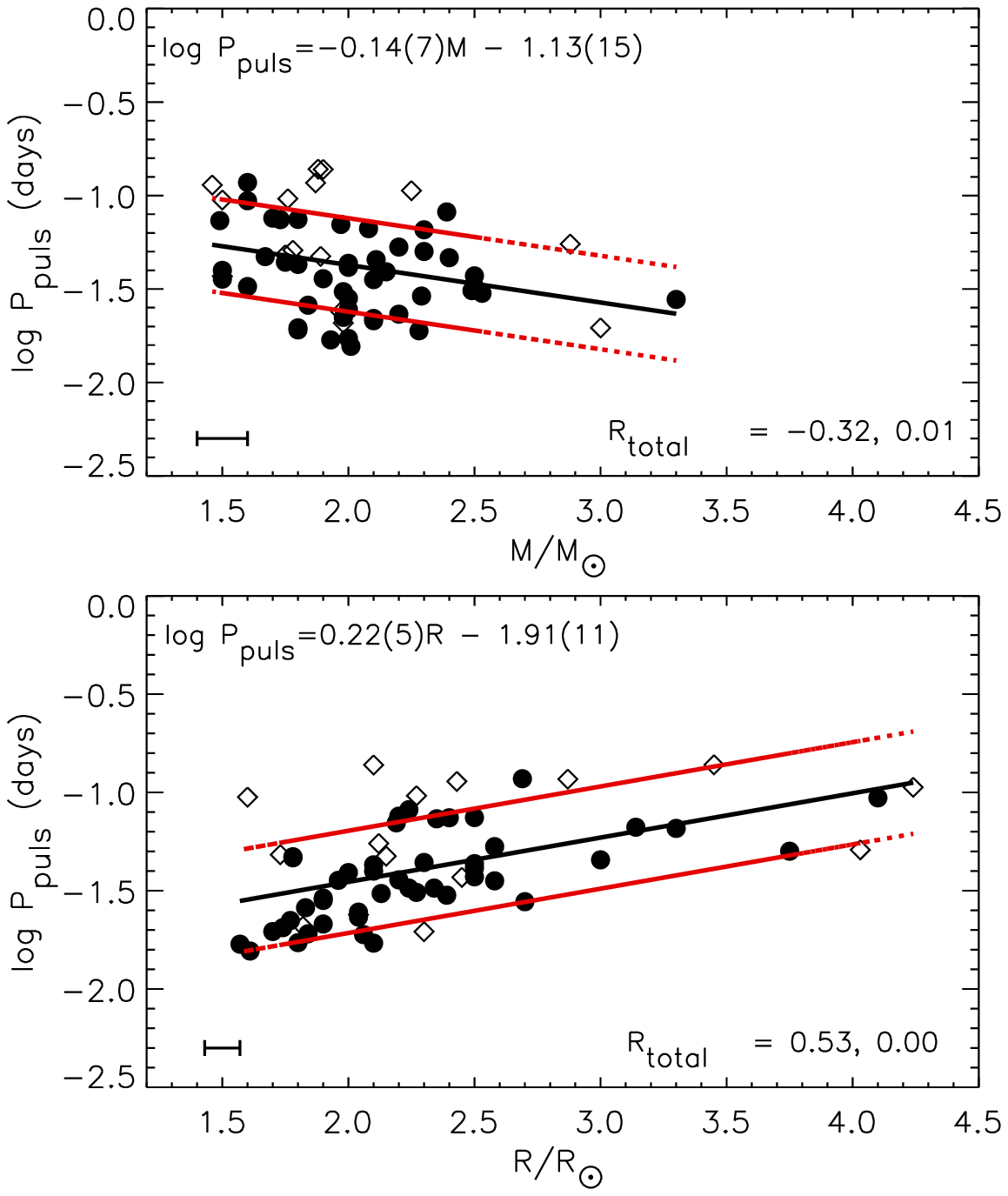}
\caption{The $M$\,--\,$\log P_{\rm puls}$ (upper panel) and $R$\,--\,$\log P_{\rm puls}$ (lower panel) correlations. 
The equation in lower panel was derived from the correlation for both detached and semi-detached systems.
The symbols, red lines, and R constant are the same as in Fig.\,\ref{figure6}.}
\label{figure9}
\end{figure}

The existence of {\logg}\,--\,$\log P_{\rm puls}$ correlation offers us other probable connections
between $M$, $R$, and $P_{\rm puls}$. Given that $\logg  \propto M/R^{2}$, a positive 
correlation for $R$\,--\,$\log P_{\rm puls}$ and a negative correlation for
$M$\,--\,$\log P_{\rm puls}$ should exist.  Hence, these were examined and the expected
correlations were obtained as demonstrated in  Fig.\,\ref{figure9}. The positive $R$\,--\,$\log P_{\rm puls}$
correlation is stronger than the negative $M$\,--\,$\log P_{\rm puls}$ correlation. Additionally,
no meaningful $M$ and $R$  correlations with $V$-band $Amp$ were detected for
all types of binaries with primary $\delta$~Sct components.  

\begin{figure}
\includegraphics[width=8.5cm,height=8.5cm, angle=0]{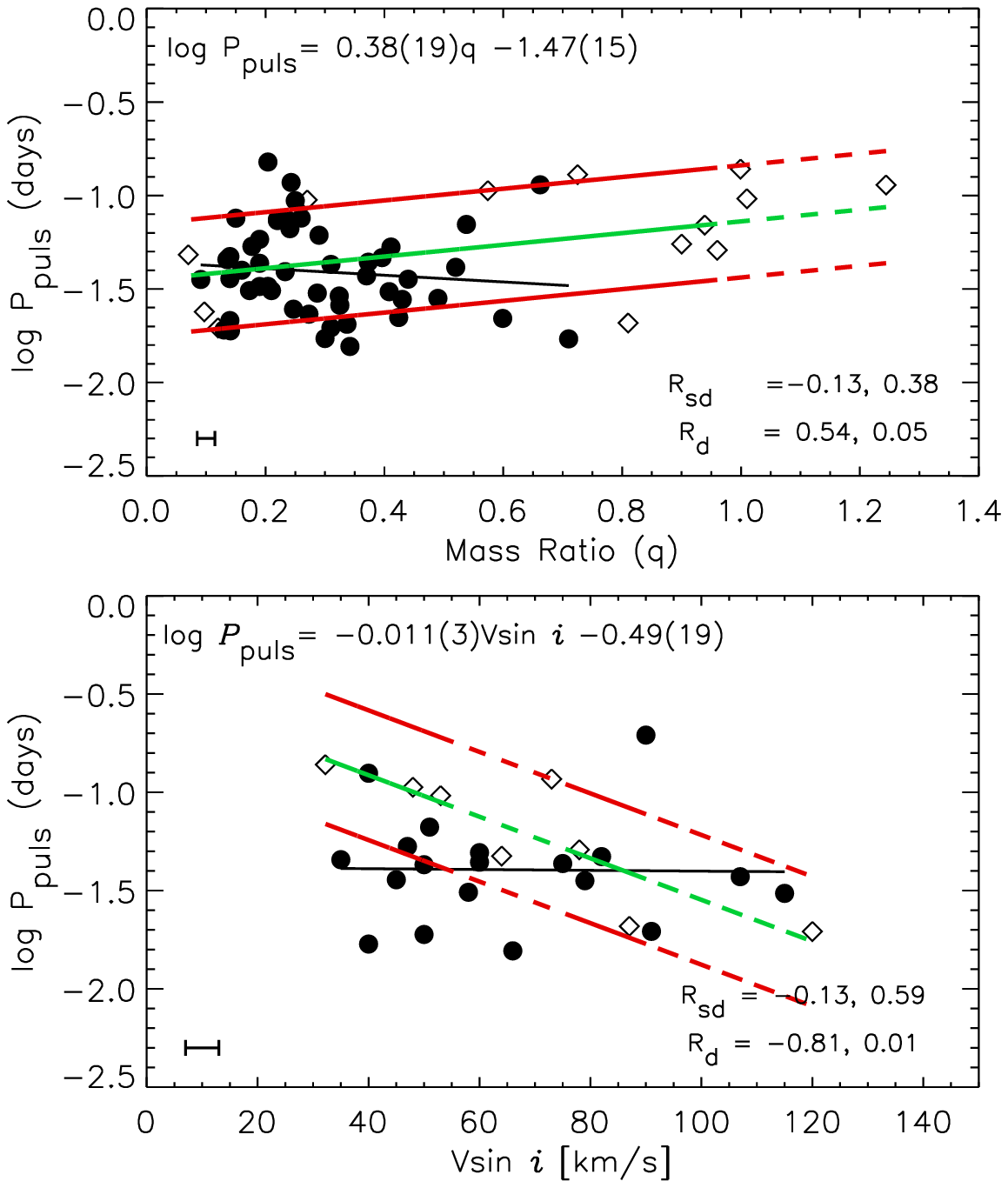}
\caption{The $q$\,--\,$\log P_{\rm puls}$ (upper panel) and {\vsini}\,--\,$\log P_{\rm puls}$ (lower panel) correlations.
Green lines illustrate the correlations for detached systems, while black lines show semi-detached systems' correlations. 
The equations in each panel were derived from the correlations of detached systems.
The symbols, red lines and R constant are the same as in Fig.\,\ref{figure6}.}
\label{figure10}
\end{figure}

According to Eq.~\ref{eq1}, theoretically a correlation between $P_{\rm puls}$ and $q$
should exist. When this relation was examined, it turned out that a correlation is present for
detached systems, although no significant correlation is found for semi-detached
systems. These are shown in the upper panel of Fig.\,\ref{figure10}. The {\vsini}\,--\,$\log P_{\rm puls}$
correlation was examined as well. This correlation is also not significant for semi-detached systems, while
there is a strong correlation between {\vsini} and $P_{\rm puls}$ for detached systems. This
relation is shown in the lower panel of  Fig.\,\ref{figure10}.

\begin{figure}
\includegraphics[width=8.5cm,height=8.5cm, angle=0]{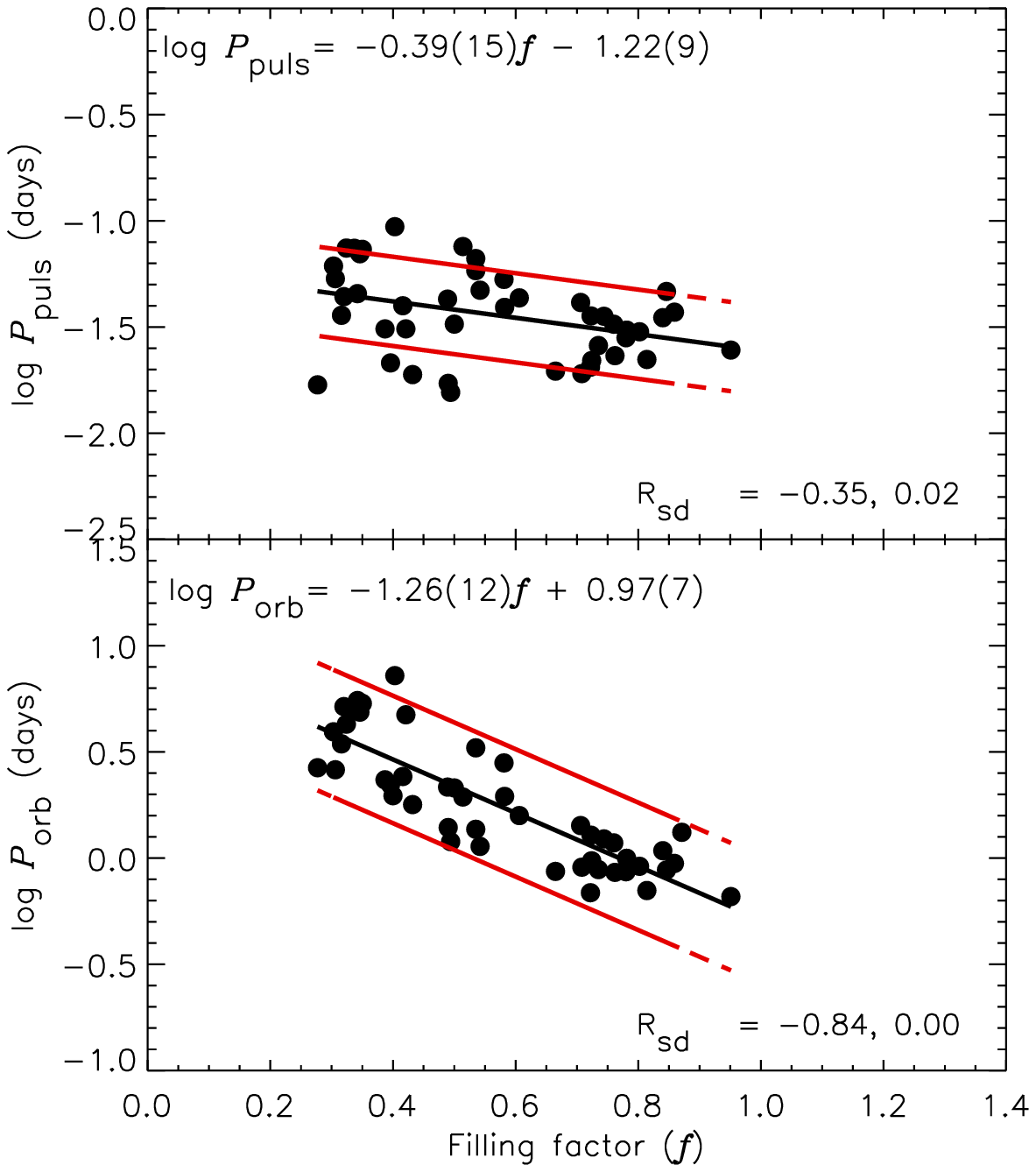}
\caption{The correlation between $P_{\rm puls}$, $P_{\rm orb}$, and $f$. 
The equations in the panels were derived from the correlations for semi-detached systems. 
No errors of $f$ values were given in the literature, hence we could not show the error bars of $f$ values. 
The symbols, red lines, and R constant are the same as in Fig.\,\ref{figure6}} 
\label{figure11}
\end{figure}

The other important factor that theoretically affects $P_{\rm puls}$ according to Eq.~\ref{eq1},
is the filling factor ($f$) of primary $\delta$~Sct components. A direct proportional
relation between $f$ and $P_{\rm puls}$ should exist. When this relation was examined for
semi-detached  systems, $f$ was found to be inversely related to $P_{\rm puls}$ as shown in the upper
panel of  Fig.\,\ref{figure11}. This result conflicts with Eq.~\ref{eq1}. We also investigated the
$f$\,--\,$P_{\rm orb}$ correlation. As shown in the lower panel of Fig.\,\ref{figure11}, $f$ regularly
decreases with increasing $P_{\rm orb}$. 

\begin{figure}
\includegraphics[width=8.5cm, height=8cm,angle=0]{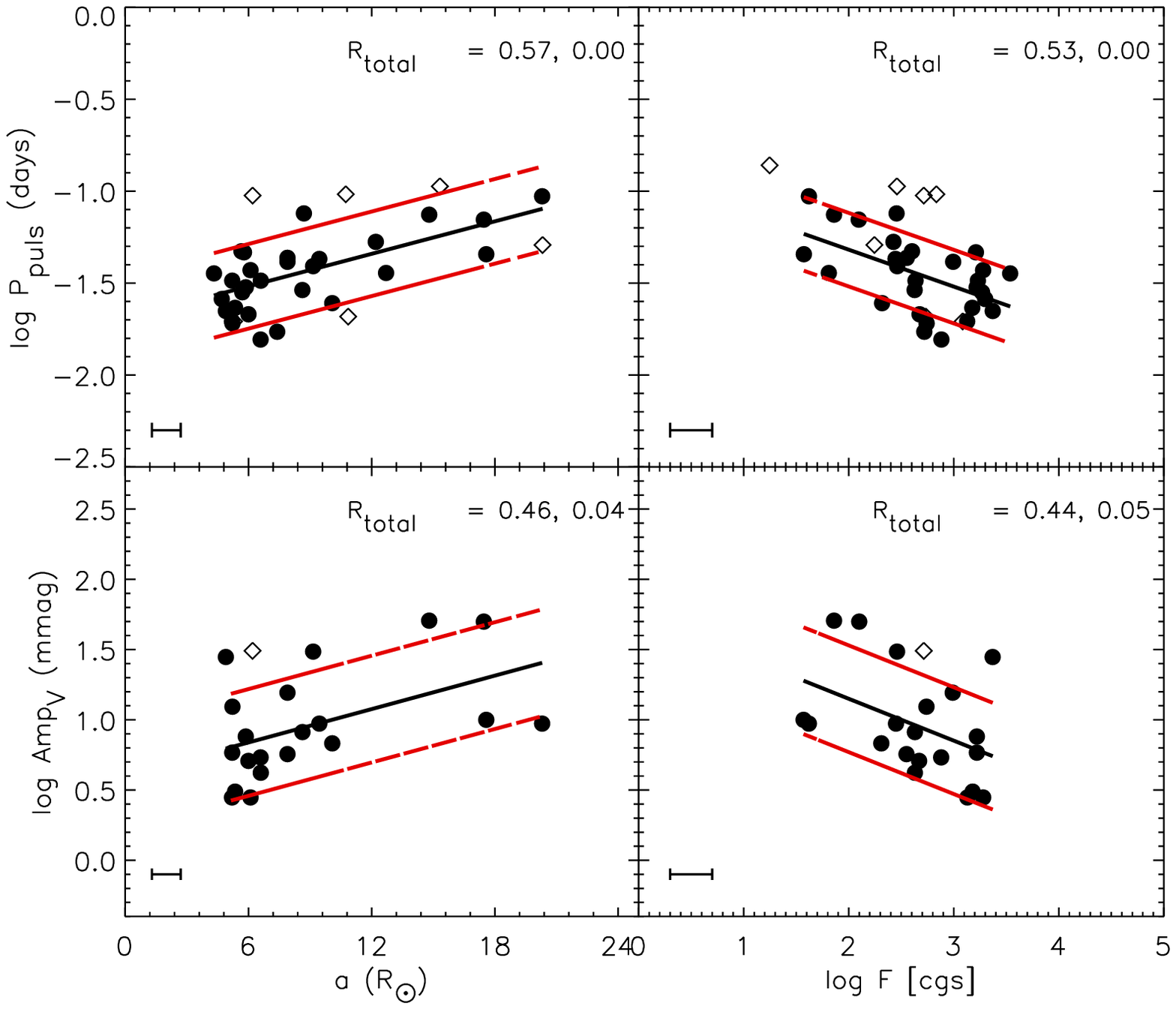}
\caption{The correlations of a and $F$ with the $P_{\rm puls}$ and V-band pulsation $Amp$ values. 
The symbols, red lines, and R constant are the same as in Fig.\,\ref{figure6}. }
\label{figure12}
\end{figure}

Additionally, we calculated the gravitational force ($F$) which is applied by the
secondary component to the primary pulsating $\delta$~Sct star. The effect of this force causes
a decrease in $P_{\rm puls}$. This result was first obtained by \citet{2006MNRAS.366.1289S}. They
found the same result as we show in the right-hand, upper panel of Fig.\,\ref{figure12}. The relation
between $F$ and $V$-band pulsation $Amp$ was examined as well and a negative correlation was found.
The relationships for these correlations were found to be:

\begin{equation} \label{eq3}
\log P_\mathrm{puls}= -0.25(6)F - 0.75(17)
\end{equation}
\begin{equation} \label{eq4}
\log Amp= -0.29(15)F + 1.76(42)
\end{equation}

In the left-hand of Fig.\,\ref{figure12} we show the correlations between orbital separation ($a$),
$P_{\rm puls}$, and $V$-band $Amp$ values. These correlations are opposite to the correlations
found for $F$ as expected, because $F \propto a^{-2}$.

\section{Discussion}

\subsection{Comparison of single and eclipsing binary member $\delta$~Sct stars}

In this section, we compare the properties of single and eclipsing binary
member $\delta$~Sct stars. All parameters  of single $\delta$~Sct stars were taken from \citet{2000A&AS..144..469R}
(R2000, hereafter). 

The $P_{\rm puls}$ values of $\delta$~Sct components in eclipsing binaries were
found between $\sim$0.016 and 0.147 days, while the $P_{\rm puls}$
range for single $\delta$~Sct stars extends to 0.288 days
(R2000). Hence, $P_{\rm puls}$ values of single $\delta$~Sct stars are significantly
longer than those in eclipsing
binaries. Additionally, highest $V$-band $Amp$ value of
single $\delta$~Sct stars is 250\,mmag (R2000), compared to only 80\,mmag
for $\delta$~Sct stars in eclipsing binaries\footnote{HADS
stars  were omitted in the comparison.}. This difference was mentioned
in the study of
\citet{2006MNRAS.370.2013S}.
Furthermore, the binarity effect was found when the average values
of $P_{\rm puls}$ of semi-detached and detached systems were compared. 
Oscillations of $\delta$~Sct stars in detached systems were found to be slower
($\sim$0.073 d$^{-1}$) than for semi-detached systems ($\sim$0.045 d$^{-1}$).
Because semi-detached systems have generally lower
$P_{\rm orb}$ values than detached systems, tidal locking is more effective
in these systems. Additionally, in semi-detached systems the secondary components
are evolved stars and they transfer mass onto the primary pulsation
components. However, no difference was found between the $V$-band pulsation $Amp$
of detached and semi-detached systems. The reason of this could also be the 
effect of mass-transfer in semi-detached systems.

The {\teff} and {\logg} of $\delta$~Sct components in eclipsing binaries were
found in the ranges of 6750\,--\,9660\,K and 3.40\,--\,4.38,
respectively. All types of eclipsing binary member $\delta$~Sct stars
have the same {\teff} ranges, although the evolved stars, which have {\logg}
values lower than 3.80, are generally detached type eclipsing binary
systems, except for QY Aql which probably has an incorrect {\logg} value.
The {\teff} of $\delta$~Sct stars is typically in the range of
6300\,--\,8600\,K \citep{2011A&A...534A.125U}. The values of $\delta$~Sct
stars in eclipsing binaries are in a good agreement with this range. However,
there are some hotter stars and the {\teff} of these stars should be
re-examined. Comparisons of {\teff} and {\logg} for single and eclipsing
binary member $\delta$~Sct stars were not made, owing to a lack of these
parameters for single $\delta$~Sct stars in R2000. 

The {\vsini} values of primary $\delta$~Sct components in eclipsing binary systems were
found between 12 and 130\,{\kms}, but extends to 285\,{\kms} for
single $\delta$~Sct stars (R2000). The average {\vsini} values for single and
binary member $\delta$~Sct stars are 90 and 64\,{\kms}, respectively. As a whole,
the single $\delta$~Sct stars rotate faster than those in eclipsing binary systems.

\subsection{Correlations}

A positive correlation between $P_{\rm puls}$ and $P_{\rm orb}$ was found for both detached and semi-detached
eclipsing binaries' primary $\delta$~Sct components. 
According to this correlation, $P_{\rm puls}$ of primary $\delta$~Sct components increase with the growing $P_{\rm orb}$. 
Growing $P_{\rm orb}$ values relate to increasing $a$ ($P_{\rm orb}$ $\varpropto$ a$^{3/2}$). 
Therefore, the effect of the secondary component on the primary pulsating component decreases with increasing 
$P_{\rm orb}$ and the pulsations of the primary $\delta$~Sct 
stars are less influenced by binarity.

\begin{figure}
\includegraphics[width=8.5cm, angle=0]{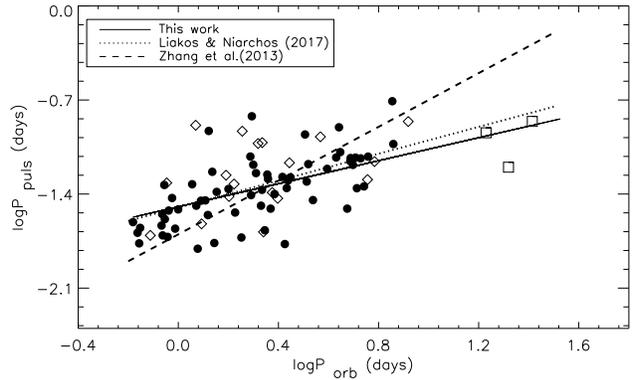}
\caption{Comparison of $\log P_{\rm orb}$\,--\,$\log P_{\rm puls}$ correlations for eclipsing binaries with a
primary $\delta$~Sct component. Square symbols illustrate the stars have $P_{\rm orb}$ $>$ 13 d.
The other symbols are the same as in Fig.\,\ref{figure6}.}
\label{figure13}
\end{figure}

The $P_{\rm puls}$\,--\,$P_{\rm orb}$  correlation was shown in the recent study of \citet{2017MNRAS.465.1181L} for all 
known $\delta$~Sct stars in binaries, including the non-eclipsing ones. They found that there is a 13-days limit in $P_{\rm orb}$ and for longer 
$P_{\rm orb}$ values binarity has less of an effect on pulsations. However, our result is different. 
In our $P_{\rm puls}$\,--\,$P_{\rm orb}$ correlation, there are detached stars (GK Dra, KIC\,3858884 and KIC\,8569819) which have $P_{\rm orb}>13$~days and 
agree with the $P_{\rm puls}$\,--\,$P_{\rm orb}$ correlation to within the 1-$\sigma$ level. The 13-days $P_{\rm orb}$ limit for the binarity effect on pulsations appears to be 
underestimated. Our results show that binarity still influences the pulsations of primary $\delta$~Sct components with $P_{\rm orb} > 13$~days.
Although \citet{2017MNRAS.465.1181L} did not include stars having $P_{\rm orb} >13$~days in their $\log P_{\rm orb}$\,--\,$\log P_{\rm puls}$ correlation, our correlation is in agreement, 
as can be seen from  Fig.\,\ref{figure13}. However, the 
theoretically calculated $\log P_{\rm orb}$\,--\,$\log P_{\rm puls}$ relationship by \citet{2013ApJ...777...77Z} is different than ours. The reason of this difference could be the negative effects 
of some parameters ($f$ and $q$ in semi-detached systems) on the pulsations, contrary to the expected positive effects of 
these parameters according to Eq.~\ref{eq1}, 
which were used to derive the theoretical $P_{\rm puls}$\,--\,$P_{\rm orb}$ 
relationship.

The $V$-band $Amp$ of primary $\delta$~Sct components in eclipsing binary systems increases
with increasing $P_{\rm orb}$. No $Amp$\,--\,$P_{\rm orb}$ correlation was
found in previous studies 
\citep{2006MNRAS.366.1289S,2012MNRAS.422.1250L,2013ApJ...777...77Z,
2017MNRAS.465.1181L}. The gravitational force applied by secondary components 
onto the surface of primary pulsation stars appears to cause a decrease in 
$Amp$.

A significant negative correlation was found between {\teff} and $P_{\rm puls}$. \citet{2011MNRAS.417..591B} 
also showed the same relation and Kahraman Ali\c{c}avu\c{s} et al. (in preparation) also found it for single $\delta$~Sct stars. 
The {\teff} and $P_{\rm puls}$ relation is an expected result. When the pulsation constant ($Q = P_{\rm puls} (\overline{\rho} / \overline{\rho}_{\sun})^{0.5}$), 
mean density ($\overline{\rho} \sim  M/R^{3}$) and the luminosity-mass relation ($L/L_{\sun} \approx M/M_{\sun}$) are 
taken into account, a negative relation between $P_{\rm puls}$ and {\teff} is found ($P_{\rm puls} \varpropto (R/R_{\sun})^{0.5}
({\teff}/ \teff_{\sun})^{-2}$). Additionally, changes in {\teff} bears on the changes in $R$, which affect the region of
He ionization which is responsible from the pulsations \citep{1980cox}.

The known negative correlation between {\logg} and $P_{\rm puls}$ was demonstrated using the data of 
newly discovered stars. The correlation shows that main-sequence $\delta$~Sct components in eclipsing 
binaries pulsate in shorter periods than evolved stars. Using the pulsation constant, mean density and surface gravity ($g \sim M/R^{2}$), a relationship between 
$P_{\rm puls}$ and $g$ can be found ($P_{\rm puls} \varpropto
g^{-0.5} R^{0.5}$). According to this rough approach, our {\logg}\,--\,$\log P_{\rm puls}$ correlation was found as expected.

\begin{figure}
\includegraphics[width=8.5cm, angle=0]{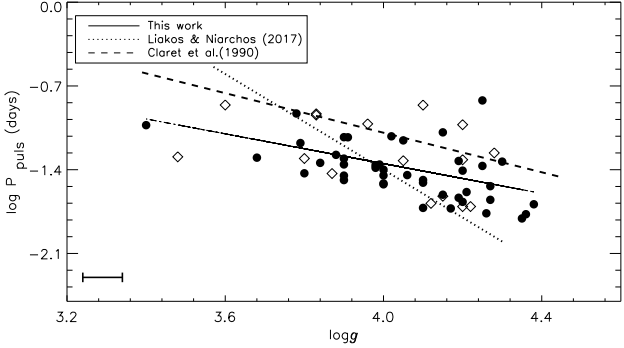}
\caption{Comparison of {\logg}\,--\,$\log P_{\rm puls}$ correlations of single and eclipsing binary member $\delta$~Sct stars. 
The symbols are the same as in Fig.\,\ref{figure6}.}
\label{figure14}
\end{figure}

The {\logg}\,--\,$\log P_{\rm puls}$ correlation was also examined by \citet{2017MNRAS.465.1181L} for 
$\delta$~Sct components in binary systems. 
Additionally, \citet{1990ASPC...11..481C} obtained the same relation for single $\delta$~Sct stars. In  Fig.\,\ref{figure14}, we 
compare the correlations of {\logg}\,--\,$\log P_{\rm puls}$ found for single and binary $\delta$~Sct stars. 
Our correlation is approximately parallel to the correlation found for single $\delta$~Sct stars, but 
there is a significant difference between our correlation and that of \citet{2017MNRAS.465.1181L}. In our study, 
we only used $\delta$~Sct stars in eclipsing binaries, whereas \citet{2017MNRAS.465.1181L} used 
all binaries containing $\delta$~Sct components. In eclipsing binaries, the {\logg} values of pulsating 
components can be derived more accurately, which is probably the reason for the difference between the two correlations.

Positive $R$\,--\,$\log P_{\rm puls}$ and negative $M$\,--\,$\log P_{\rm puls}$ correlations were obtained, as expected.
From the {\logg}\,--\,$\log P_{\rm puls}$ correlation we know that both $M$ and $R$ values have effect on pulsation.
Therefore, combining both equations we obtain:
\begin{equation} \label{eq5}
\log P_\mathrm{puls}= 0.11 (R/R_{\sun}) - 0.07 (M/M_{\sun}) - 1.52
\end{equation}
A similar equation was found for Cepheid stars by \citet{1965ApJ...142.1072F}. As can be seen from the equation, 
$P_{\rm puls}$ is more influenced by changes in $R$ than changes in $M$. 
In  Fig.\,\ref{figure9}, the weak effect of $M$ and the stronger effect of $R$ on the pulsation of primary $\delta$~Sct components 
can be seen.

We found that the binary mass ratio ($q$) has no significant effect on $P_{\rm puls}$
of primary $\delta$~Sct components in semi-detached systems, although there is a correlation between $q$ and 
$P_{\rm puls}$ for primary $\delta$~Sct components in detached systems. According to Eq.~\ref{eq1}, $P_{\rm puls}$ 
should be directly proportional to $q$. The lack of any correlation in semi-detached systems might be due to the lack of systems with $q >0.5$.

The variation of $P_{\rm puls}$ with {\vsini} was also found only for detached systems. The $P_{\rm puls}$ 
decreases with increasing \vsini. Since semi-detached systems 
are generally close binaries, rotation synchronisation is present. Therefore, 
owing to the $P_{\rm puls}$\,--\,$P_{\rm orb}$ correlation, we expected to find a correlation between $P_{\rm puls}$ and {\vsini} 
in the semi-detached systems. However, mass-transfer in 
these systems is very effective 
and this changes the $q$ and $R$ of the primary $\delta$~Sct components, and these affect the rotation and angular momentum. 
The altered angular momentum also results in a change of $P_{\rm orb}$ which can change the rotation
($P_{\rm orb} = \sqrt{2\pi}/\omega$). All these effects can be the reason why we did not find a {\vsini}\,--\,$P_{\rm puls}$ 
correlation for semi-detached systems. A correlation between $P_{\rm puls}$ and {\vsini} was also found by \citet{2013A&A...556A..52T} using the {\vsini} values of 
some $\delta$~Sct stars taken from \citet{2011A&A...534A.125U} and R2000. In their work 
they found a weak $P_{\rm puls}$\,--\,{\vsini} relation, but, contrary to our results, with $P_{\rm puls}$ increasing with declining {\vsini}. The rotation of stars causes changes in their
stellar structure, hence the reason why $P_{\rm puls}$ can be different for different values of {\vsini} \citep{1998A&A...334..911S}.

According to Eq.~\ref{eq1},  
$f$ should be directly proportional to $P_{\rm puls}$. However, in our study, we have obtained the opposite result. 
When the correlation between $P_{\rm orb}$ and $f$ was examined, we noticed that $f$ increases with 
decreasing $P_{\rm orb}$. The gravitational force applied on the primary pulsating component grows with increasing $f$ value. 
Thus, we can say $P_{\rm orb}$ has a significant effect on $f$ and we, therefore, 
obtained a negative relation between $P_{\rm puls}$ and $f$ instead of a positive correlation.
Additionally, we found that the strength of $F$ applied by the secondary component to the primary pulsation star 
affects $P_{\rm puls}$ and $Amp$ in a negative way. The same correlation between $F$ and $P_{\rm puls}$ was also obtained by 
\citet{2006MNRAS.366.1289S}.\\

\subsection{Positions in HR Diagram}

\begin{figure*}
\includegraphics[width=13cm, angle=0]{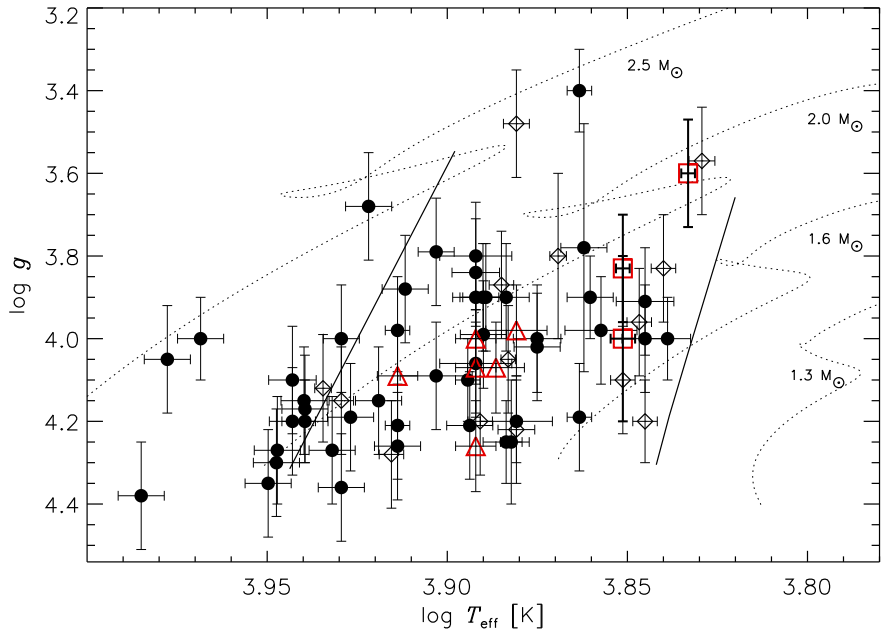}
\caption{Positions of analysed and collected $\delta$~Sct stars in eclipsing binary systems. The symbols are the same as in Fig.\,\ref{figure6}.  
Solid lines represent the theoretical instability strips of $\delta$\,Sct stars \citep{2005A&A...435..927D}. Triangle and 
square symbols illustrate the stars analysed spectroscopically in this study and the stars have $P_{\rm orb}$ $>$ 13 d, respectively. The evolutionary tracks were taken from \citet{2016MNRAS.458.2307K}}
\label{figure15}
\end{figure*}

The positions of the analysed primary $\delta$~Sct components in this study and the other 
primary $\delta$~Sct components given in the updated list in the Hertzsprung-Russell (HR) Diagram are shown in
Fig.\,\ref{figure15}. Our analysed $\delta$~Sct components and the $\delta$~Sct components given in the revised list are 
located in the $\delta$~Sct instability strip. However, there are a few stars (RR Lep, V2365 Oph, VV UMa and V346 Cyg) 
placed beyond the blue edge of $\delta$~Sct instability 
strip. The {\teff} and {\logg} values of these stars were taken from literature spectral classification and photometric
analyses (see Table\,A1 for references). Therefore, these stars should be re-analysed with new data.

In our study, the primary $\delta$~Sct components in eclipsing binaries were mostly located inside the theoretical $\delta$~Sct
instability strip to within the error bars. However, in the study of \citet{2017MNRAS.465.1181L}, 
there are more stars located beyond the blue border of $\delta$~Sct instability strip compared to our results. 
\citet{2017MNRAS.465.1181L} showed positions of $\delta$~Sct stars in all binaries, whereas we only showed the positions of 
$\delta$~Sct in eclipsing binaries. The fundamental parameters of stars can be obtained more accurately in the eclipsing 
binary case. Probably because of this reason \citet{2017MNRAS.465.1181L} found more stars located beyond the 
blue edge of $\delta$~Sct instability strip.

\section{Conclusions}

In this study, we present an updated list of $\delta$~Sct stars in eclipsing binaries and the spectroscopic 
analysis of six of $\delta$~Sct components in eclipsing binary systems.

In the spectroscopic analysis of six primary $\delta$~Sct components in eclipsing binaries, we obtained the 
spectral classification, \teff, \vsini, and $[m/H]$ of the stars. XX Cep was found to be 
metal-rich, while others have approximately solar metallicity.

In the updated list of $\delta$~Sct components in eclipsing binaries, we collected the atmospheric and orbital parameters of 
the primary $\delta$~Sct components. We examined the properties of the primary $\delta$~Sct components 
and compared them with the properties of single $\delta$~Sct stars. 
\citet{2017MNRAS.465.1181L} stated that the single and binary member $\delta$~Sct stars have similar 
pulsational behaviour. However, when $P_{\rm puls}$ and $V$-band $Amp$ of single and eclipsing binary member 
$\delta$~Sct stars were compared, we found that eclipsing binary member $\delta$~Sct stars oscillate 
with shorter $P_{\rm puls}$ and lower $Amp$ comparing to single ones. These differences in pulsation 
quantities of single and binary $\delta$~Sct stars are thought to be caused by the effects of gravitational force applied 
by the secondary component on the primary and mass-transfer in these binaries. Additionally, binarity effects were also
found when $P_{\rm puls}$ of detached 
and semi-detached member $\delta$~Sct stars were compared. We showed that $\delta$~Sct 
stars in detached systems pulsate in longer periods. 

The {\vsini} of single and eclipsing binary member $\delta$~Sct stars was also compared. We found that, on average, single $\delta$~Sct stars 
rotate faster than those in eclipsing binary systems.

We examined the relations between the orbital and atmospheric parameters of primary $\delta$~Sct components. 
Firstly, the known $P_{\rm puls}$\,--\,$P_{\rm orb}$ correlation was checked and we obtained that $P_{\rm puls}$ increases with increasing $P_{\rm orb}$. 
\citet{2017MNRAS.465.1181L} found that binarity does not have a significant effect on pulsation if $P_{\rm orb}$ $\geqslant$ 13 days.
However, we showed that the $P_{\rm puls}$\,--\,$P_{\rm orb}$ correlation is still significant 
even if $P_{\rm orb}$ is 26 days. Therefore, it appears that the 13-days limit for the binarity effect is too low.
When our $P_{\rm puls}$\,--\,$P_{\rm orb}$ correlation was compared with the previously found correlations 
we obtained similar trends except for the theoretically calculated relationship  of \citet{2013ApJ...777...77Z}. The difference between the theoretical relation and our 
correlation is caused by some parameters ($f$, $q$) having adverse effects on pulsation, whereas these parameters were found to be directly proportional 
to pulsation in the theory. We also found that $V$-band $Amp$ of primary $\delta$~Sct components increases with increasing $P_{\rm orb}$.

Significant negative relations between $P_{\rm puls}$ and atmospheric parameters {\teff} and {\logg} were found. 
The {\logg}\,--\,$\log P_{\rm puls}$ correlation was already known, however the $P_{\rm puls}$\,--\,{\teff} correlation
for the primary $\delta$~Sct
components was shown the first time. The {\logg}\,--\,$\log P_{\rm puls}$ correlation was compared with those in the literature. We find that our 
correlation is almost in agreement with that found for single $\delta$~Sct stars. However, the correlation found by \citet{2017MNRAS.465.1181L} is 
incompatible with ours. 

A positive $R$\,--\,$\log P_{\rm puls}$ and a negative $M$\,--\,$\log P_{\rm puls}$ correlations were found. 
As both parameters influence the pulsation, we gave a new equation for $P_{\rm puls}$ in terms of $R$ and $M$ (Eq.~\ref{eq5}). Additionally, we showed that increasing $q$ caused increasing in $P_{\rm puls}$ 
for detached systems, while $q$ has no effect on $P_{\rm puls}$ in semi-detached systems. According to theory $P_{\rm puls}$ should be directly 
proportional to $q$. The relationship 
between $P_{\rm puls}$ and {\vsini} of primary $\delta$~Sct components was also examined. No relationship was obtained for
semi-detached systems. However, for detached systems, $P_{\rm puls}$ of the primary $\delta$~Sct components decreases with
increasing \vsini. The suggested positive $f$ and $P_{\rm puls}$ correlation by \citet{2013ApJ...777...77Z} 
was also checked. However, we found that $P_{\rm puls}$ is inversely proportional to $f$. 
When the relationship between $P_{\rm orb}$ and $f$ was checked, we also obtained a negative correlation. 
Components in binaries come closer to each other with decreasing $P_{\rm orb}$ and  the Roche lobes of the components become smaller, 
therefore $f$ increases with decreasing $P_{\rm orb}$. This effect is rather dominant in binaries. Hence, we still see this effect in the $f$\,--\,$P_{\rm puls}$ relationship. Therefore, a negative correlation 
between these parameters was obtained contrary to suggested relation. Additionally, we 
found that the gravitational force applied by the secondary components onto the primary $\delta$~Sct components changes $P_{\rm puls}$ and $Amp$ of 
$\delta$~Sct stars.

The positions of the primary $\delta$~Sct components in the $\log
{\teff}$\,-\,{\logg} diagram were shown. The primary $\delta$~Sct components in detached and semi-detached systems
are located inside the $\delta$~Sct instability 
strip. However, there are some semi-detached member $\delta$~Sct components located beyond the blue edge of $\delta$~Sct instability 
strip, but the {\teff} and {\logg} of these stars may not be reliable.

In this study, we show the importance of $\delta$~Sct components in eclipsing binaries. The differences between the single and 
binary member $\delta$~Sct stars were emphasised. The effects of the fundamental and orbital parameters on pulsation and the correlations 
between the pulsation quantities and some fundamental parameters were given. These relationships allow us to infer the initial values of  
the fundamental parameters of pulsating $\delta$~Sct components. This is important for the theoretical examination of pulsating stars and 
understanding the internal structures and evolutionary statuses of stars. 
Additionally, utilizing the found $P_{\rm puls}$\,--\,$P_{\rm orb}$ correlation, the lower frequencies in 
$\delta$~Sct stars can be examined to see if they are related to binarity.  

\section*{Acknowledgments}
The authors would like to thank the reviewer for useful comments
and suggestions that helped to improve the publication. 
This work has been partly supported by the Scientific and Technological Research Council of Turkey (TUBITAK) grant numbers 2214-A and 2211-C. 
We thank \c{C}anakkale Onsekiz Mart University Research Foundation (Project No. FDK\,-\,2016\,-\,861) for supporting this study.
This article is a part of the PhD thesis of FKA. JK thanks to the grant 16-01116S (GA\v{C}R).
We thank Dr. G. Catanzaro for putting the code for Balmer lines analysis at our disposal. 
We are grateful to Dr. D. Shulyak for putting the code for calculating ${\it E(B-V)}$ at our disposal. 
We thank Dr. J. Ostrowski for helping us for the evolution tracks.  
This work has made use of data from 
the European Space Agency (ESA) mission Gaia (http://www.cosmos.esa.int/gaia), processed by the Gaia Data Processing 
and Analysis Consortium (DPAC, http://www.cosmos.esa.int/web/gaia/dpac/consortium). Funding for the DPAC has been 
provided by national institutions, in particular the institutions participating in the Gaia Multilateral Agreement. 
This research has made use of the SIMBAD data base, operated at CDS, 
Strasbourq, France.

\appendix
\setcounter{table}{0}
\begin{landscape}

\begin{table}
 \begin{scriptsize}

  \caption{The list of $\delta$~Sct stars in eclipsing binaries. The fundamental, orbital, and atmospheric parameters of primary $\delta$~Sct and
  secondary components. The subscripts, p and s represents primary and secondary components, respectively.}
  \label{delsctecl}
  \begin{tabular}{llllllllllllll}
  \hline
HD      & Name			  & V    & Spectral  & Parallaxes& Type & $P_{\rm orb}$ &$P_{\rm puls}$&Amp$_{V}$&Amp$_{B}$&\teff$_{p}$&\teff$_{s}$ & \logg$_{p}$& \vsini$_{p}$ \\
        &                         & (mag)&  Type     & (mas)     &      &         (days)   & (days)   & (mmag)  & (mmag)  & (K)       & (K)        &           & (km\,s$^{-1}$) \\
\hline
354963	& QY Aql		  &11.89 &F0	     & 1.74 (87) &  SD  &    7.2296 & 0.0938&  9.4 (2)   &11.8 (2)& 7300       &4244 (122)  &3.4 (10)	 &	      \\
193740	& XZ Aql		  &10.18 &A2/3 II/III& 2.03 (44) &  SD  &    2.1392 & 0.0326&  6.8 (2)   &8.6 (2) & 8770 (150) &4720 (150)  &4.10 (03)   &	   	  \\
	& V729 Aql	          &13.76 &	     &	 	 &  SD  &    1.2819 & 0.0357&  4.2 (4)   &	  & 6900       &4300 (175)  &4.00 (10)   &	   	  \\
	& V1464 Aql		  &08.68 &A2V	     & 4.12 (48) &  SD  &    0.6978 & 0.0171&  24.0 (3)  &30.0 (2)&7420 (192)  &6232 (161)  &	 	 &	   	  \\
  	& CZ Aqr		  &11.10 &A5	     &  	 &  SD  &    0.8627 & 0.0282&	         &3.7 (5) & 8200       &5650 (12)   &4.21	 &  42        \\
211705	& DY Aqr		  &10.49 &A1/2 III   & 1.83 (75) &  SD  &    2.1597 & 0.0428&  9.4	 &	  & 7625 (125) &3800 (200)  &4.25 (25)   &  50 (10)   \\
203069	& RY Aqr		  & 9.25 &A7V	     & 6.48 (28) &  SD  &	    &	    &	 	 &	  & 7650       &4520 (122)  & 4.25 (60)  &	   	  \\
	& V551 Aur		  &14.27 &F	     &	 	 &  D	&    1.1732 & 0.1294&  19.1 (3)  &15.7 (3)&7000	       &6085 (34)   &	 	 &	   	  \\	
	& EW Boo		  &10.26 &A0	     & 2.23 (26) &  SD  &    0.9063 & 0.0191&  12.4 (2)  &14.4 (2)& 7840       &4515 (35)   &4.1 (10)	 &	      \\
	& YY Boo		  &11.58 &A4+K4IV    & 1.17 (23) &  SD  &    3.9330 & 0.0613&  79.2 (2)  &116.8(2)&	       &4650 (10)   &	 	 &	      \\
	& Y Cam		  	  &10.60 &A9IV+K1IV  & 0.82 (30) &  SD  &    3.3058 & 0.0665&  12.2	 &	  & 8000 (250) &4629 (150)  &3.79	 &  51 (4)    \\
194168	& TY Cap		  &10.36 &A5 III     & 1.95 (55) &  SD  &    1.4235 & 0.0413&  15.6 (7)  &18.5 (7)& 8200       &4194 (30)   &3.98	 &	      \\
	& AB Cas		  &10.32 &A3 V       & 2.91 (22) &  SD  &    1.3669 & 0.0583&  19.6 (9)  &22.2 (1)& 8000       &4729 (24)   &	 	 &	      \\
	& IV Cas		  &11.34 &A2	     & 1.06 (40) &  SD  &    0.9985 & 0.0306&	         &3.4 (2) & 8500 (250) &5193 (7)    &4.0 (50)	 &  115 (5)   \\
017138	& RZ Cas		  & 6.26 &A3 V       & 14.99 (34)&  SD  &    1.1952 & 0.0156&  5.4 (3)   &2.7 (3) & 8907 (15)  &4797 (20)   &4.35	 &   66 (1)   \\
	& V389 Cas		  &11.09 &	     &	 	 &  D	&    2.4948 & 0.0370&  9.2 (4)   &	  & 7673 (31)  &4438 (22)   &3.87	 &	   	  \\
	& V1264 Cen		  &11.95 &A7V	     & 0.97 (39) &  SD  &    5.3505 & 0.0734&  350.0	 &	  & 7500       &4200	    &4.00	 &	      \\	
222217	& XX Cep		  & 9.18 &A7V	     & 3.17 (23) &  SD  &    2.3374 & 0.0310&  2.6 (2)   &2.9 (2) & 8000 (250) &4280 (36)   &4.09	 &   50       \\
	& WY Cet		  & 9.28 &F0V	     & 4.64 (73) &  SD  &    1.9396 & 0.0757&	         &7.7 (3) & 7500       &4347 (7)    &4.02	 &	      \\
075747	& RS Cha		  & 6.07 &A7V	     &	 	 &  D	&    1.6699 & 0.0473&		 &	  & 7640 (76)  &7230 (72)   &4.05	 &   64 (6)   \\
057167	& R Cma			  & 5.70 &F2 III/IV  & 23.3 (59) &  SD  &    1.1359 & 0.0471&		 &	  & 7300       &4350	    &4.19	 &  82 (3)    \\
	& UW Cyg		  &10.86 &A7/A6 IV   & 1.60 (30) &  SD  &    3.4507 & 0.0359&	         &1.9 (2) & 7800 (250) &4347 (4)    &4.06	 &   45       \\
	& V346 Cyg		  &12.22 &A5	     & 1.11 (22) &  SD  &    2.7433 & 0.0502&	         &30.00   & 8353       &6620	    &3.68	 &	      \\
	& V469 Cyg		  &12.33 &B8+F0      &	 	 &  SD  &    1.3124 & 0.0278&  20.0	 &	  &	       & 	    &4.13	 &	      \\
099612	& AK Crt		  &11.28 &A5/9 II/III& 1.81 (48) &  D	&    2.7788 & 0.0680&  8-35	 &	  &	       & 	    &	 	 &	      \\	
	& BW Del		  &11.28 &F2         &	 	 &  SD  &    2.4231 & 0.0398&  1.8 (2)   &2.9 (2) & 7000       &4061 (30)   &4.00	 &	      \\
152028	& GK Dra	          & 8.77 &F0	     & 3.03 (22) &  D	&    16.960 & 0.1138&		 &	  & 7100 (70)  &6878 (57)   &3.83 (03)   &	      \\
	& GQ Dra		  &	 &	     &	 	 &  SD  &    0.7659 & 0.0335&		 &	  &	       & 	    &	 	 &	   	  \\
172022	& HL Dra		  & 7.36 &A6IV       & 6.24 (24) &  SD  &    0.9443 & 0.0372&  2.8 (3)   &3.0 (2) & 7800 (250) &5074 (8)    &3.80	 &   88       \\
173977	& HN Dra		  & 8.07 &F2	     & 3.88 (23) &  D	&    1.8008 & 0.1169&  7.6	 &	  & 6918       &6309	    &3.83	 &   73       \\
187708	& HZ Dra		  & 8.14 &A8/A7 V    & 4.89 (29) &  D	&    0.7729 & 0.0196&	         &4.0 (4) & 7600 (250) &5015 (68)   &4.22	 &  120       \\
	& OO Dra		  &11.39 &	     & 1.54 (29) &  D	&    1.2384 & 0.0239&  4.2 (2)   &4.9 (3) & 8500       &6452 (8)    &4.15	 &	      \\
238811	& SX Dra		  &10.40 &A7V+ K7IV  & 1.16 (27) &  SD  &    5.1696 & 0.0440&  23.6 (3)  &34.6 (4)& 7762       &4638 (200)  &3.99	 &	      \\
139319	& TW Dra		  & 7.46 &A5+K0III   & 5.90 (24) &  SD  &    2.8069 & 0.0530&	         &10.00   & 8160 (15)  &4538 (11)   &3.88 (02)   &  47 (1)    \\
	& TZ Dra		  & 9.32 &A7V	     & 3.96 (25) &  SD  &    0.8660 & 0.0196&  2.8 (2)   &3.7 (2) & 7600 (250) &5088 (55)   &4.20 (10)   &    80      \\
021985	& AS Eri		  & 8.30 &A1V	     & 5.06 (51) &  SD  &    2.6641 & 0.0169&		 &	  & 8500       &4790	    &4.35	 &    40      \\
	& TZ Eri		  & 9.61 &A5	     & 3.27 (34) &  SD  &    2.6061 & 0.0534&  7.3       &8.30    & 9307 (20)  &4562	    &	 	 &	      \\
336759	& BO Her		  &11.14 &A7	     & 1.51 (30) &  SD  &    4.2728 & 0.0745&  50.8 (3)  &68.0 (3)& 7800       &4344 (68)   &3.90 (10)   &	      \\
	& CT Her		  &11.32 &A3V	     & 0.83 (46) &  SD  &    1.7864 & 0.0189&	         &3.3 (1) & 8700       &4651 (7)    &4.17 (02)   &    50      \\
	& EF Her		  &11.53 &A0	     & 1.14 (26) &  SD  &    4.7292 & 0.0310&  51.0      &69.0    & 9327       &4767	    &	 	 &	      \\
151973	& LT Her		  &10.55 &A1	     & 1.71 (81) &  SD  &	    &	    &	 	 &	  & 9400       &5063 (25)   &	 	 &	   	  \\	
	& TU Her		  &11.14 &A5	     & 1.84 (22) &  SD  &    2.2669 & 0.0556&  9-10	 &	  &	       & 	    &	 	 &	      \\
	& V948 Her		  & 8.91 &F2	     & 6.15 (26) &  D	&    2.0831 & 0.0947&  31.0	 &	  & 7000       &4310 (63)   &4.20 (10)   &	      \\
	& AI Hya		  & 9.35 &F2m+F0V    & 1.88 (35) &  D	&    8.2897 & 0.1380&  20.0	 &	  & 7100       &6750	    &4.10	 &	      \\
078014	& RX Hya		  & 9.56 &A5 III     & 3.80 (33) &  SD  &    2.2817 & 0.0516&  7.0	 &	  &	       & 	    &	 	 &	      \\
	& AU Lac		  &11.81 &A5	     & 1.60 (20) &  SD  &    1.3926 & 0.0172&	         &5.00    & 8200       &3784 (15)   &4.26	 &	      \\
	& WY Leo		  &10.89 &A2	     & 1.51 (49) &  SD  &    4.9858 & 0.0656&  11.0 (1)  &	  &	       & 	    &3.79	 &	      \\
	& Y Leo		  	  &10.07 &A3V	     & 2.50 (26) &  SD  &    1.6861 & 0.0290&  8.12 (15) &	  & 8855       &4276 (23)   &4.27	 &	      \\
033789	& RR Lep		  &10.14 &A4 III     & 2.20 (35) &  SD  &    0.9154 & 0.0300&  7.6 (4)   &9.6 (4) & 9300       &4904 (106)  &4.00 (10)   &	      \\
	& CL Lyn		  & 9.77 &A8 IV      & 2.82 (25) &  SD  &    1.5861 & 0.0434&  5.7 (4)   &7.3 (3) & 7200 (250) &4948 (14)   &3.98	 &   75       \\
198103	& VY Mic		  & 9.54 &A4 III/IV  & 1.66 (36) &  SD  &    4.4364 & 0.0817&  19.4 (2)  &	  & 8705       &5301	    &4.15	 &	      \\
	& V577 Oph		  &11.19 &A	     & 1.29 (26) &  D	&    6.0791 & 0.0695&  57.8	 &	  &	       & 	    &	 	 &	      \\
155002	& V2365 Oph		  & 8.86 &A2	     & 3.54 (29) &  SD  &    4.8656 & 0.0700&  50.0	 &	  & 9500       &6400 (27)   &4.05	 &	      \\
293808	& FL Ori		  &11.42 &A2	     & 2.05 (40) &  D	&    1.5510 & 0.0550&		 &	  & 8232       &5243	    &4.28	 &	      \\
248406	& FR Ori		  &10.64 &A7	     & 2.53 (86) &  SD  &    0.8832 & 0.0259&  5.8	 &	  & 7830       &4583 (10)   &4.21	 &	   	  \\
252973	& V392 Ori		  &10.49 &A5V	     & 2.56 (25) &  SD  &    0.6593 & 0.0246&		 &	  & 8300       &5065 (11)   &4.15	 &	   	  \\
	& MX Pav		  &11.35 &A5+K3IV    & 1.56 (38) &  SD  &    5.7308 & 0.0756&  76.9 (3)  &	  &	       & 	    &	 	 &	      \\
	& BG Peg		  &11.35 &A2	     &	 	 &  SD  &    1.9527 & 0.0391&  30.6 (5)  &36 (6)  & 8770       &5155 (200)  &4.20	 &	      \\
275604	& AB Per		  & 9.72 &F0	     &	 	 &  SD  &    7.1603 & 0.1954&		 &	  &	       &  	    &	 	 &	      \\
	& IU Per		  &10.56 &A4	     & 1.62 (45) &  SD  &    0.8570 & 0.0232&  3.08 (07) &	  & 8450       &4900 (250)  &4.29	 &	      \\
	& AO Ser		  &11.04 &A2	     & 2.19 (41) &  SD  &    0.8793 & 0.0465&	         &20.00   & 8860       &4547 (512)  &4.30	 &	      \\
\hline
\end{tabular}
 \end{scriptsize}
\end{table}
 \end{landscape}

\setcounter{table}{0}
\begin{landscape}

\begin{table}
 \begin{scriptsize}

  \caption{Continuation.}
  \begin{tabular}{llllllllllllll}
  \hline
HD      & Name			  & V    & Spectral  & parallaxes& Type & $P_{\rm orb}$ &$P_{\rm puls}$&Amp$_{V}$&Amp$_{B}$&\teff$_{p}$&\teff$_{s}$ & \logg$_{p}$& \vsini$_{p}$\\
        &                         & (mag)&  Type     & (mas)     &      &  (days)   & (days)   & (mmag)  & (mmag)  & (K)       & (K)        &            & (km\,s$^{-1}$)\\
\hline	
	& UZ Sge		  &11.40 &A0	     &	 	 &  SD  &    2.2157 & 0.0214&		 &	  & 8700       &4586 (60)   &4.20 (10)   &	   	  \\
	& AC Tau		  &11.09 &A8	     & 1.65 (43) &  SD  &    2.0434 & 0.0570&  6.0	 &	  &	       & 	    &	 	 &	      \\
	& IZ Tel		  &12.20 &A8+G8 IV   & 0.52 (26) &  SD  &    4.8802 & 0.0738&  45.9 (4)  &	  &	       & 	    &	 	 &	      \\
12211	& X Tri			  & 9.00 &A7V	     & 4.85 (22) &  SD  &    0.9715 & 0.0220&  20.0	 &	  & 8600       &5188 (4)    &	 	 &	   	  \\	
115268	& IO Uma		  & 8.21 &A3	     & 1.05 (43) &  SD  &    5.5202 & 0.0454&  10 (2)    &13.0 (2)& 7800 (150) &4260 (30)   &3.84 (05)   &   35 (2)	  \\
	& VV Uma		  &10.28 &A2V	     & 2.45 (45) &  SD  &    0.6874 & 0.0205&  28.0 (1)  &	  & 9660 (30)  &5579 (20)   &4.38	 &	   	  \\
	& AW Vel		  &10.70 &A7	     & 1.69 (41) &  SD  &    1.9925 & 0.0658&  58.0 (1)  &	  &	       & 	    &	 	 &	   	  \\	
	& BF Vel		  &10.62 &A3	     & 1.92 (30) &  SD  &    0.7040 & 0.0223&	         &26.0 (2)& 8550       &4955 (4)    &4.27	 &	   	  \\
	& CoRot 105906206	  &12.21 &	     & 0.96 (25) &  D	&    3.6946 & 0.1062&		 &	  & 6750 (150) &6152 (162)  &3.57	 &   48 (1)	  \\
172189	& GSC 455-1084		  & 8.73 &A6V-A7V    &	 	 &  D	&    5.7017 & 0.0510&		 &	  & 7600 (150) &8100 (150)  &3.48	 &   78 (3)   \\
232486	& GSC 3671-1094		  & 9.64 &A5	     & 3.07 (25) &  D	&    2.3723 & 0.0409&  20.0	 &	  &	       & 	    &	 	 &	      \\
        & GSC 3889-202		  &10.39 &A7 V-IV    & 1.27 (24) &  SD  &    2.7107 & 0.0441&  50.0      &70.0    & 7750       &4500	    &3.90	 &    60      \\
	& GSC 4293-432		  &10.56 &A7+K3      &	 	 &  SD  &    4.3844 & 0.1250&  35.0      &40.0    & 7750       &4300	    &		 &    40      \\
	& GSC 4588-883		  &11.31 &A9 IV+K4III& 0.94 (48) &  SD  &    3.2586 & 0.0493&		 &	  & 7650       &4100	    &3.90	 &    60      \\
062571	& GSC 4843-2140		  & 8.83 &F0-F2      &	 	 &  SD  &    3.2087 & 0.1141&  41.7	 &	  & 7762       &5719 (150)  &	 	 &	      \\
220687	& GSC 5825-1038		  & 9.60 &A2 III     & 2.31 (42) &  D	&    1.5943 & 0.0382&  12.8 (14) &	  &	       & 	    &	 	 &	      \\
	& KIC 3858884		  & 9.28 &F5	     & 1.78 (22) &  D	&    25.952 & 0.1383&		 &	  & 6810 (70)  &6890 (80)   &3.60	 &   32 (2)	  \\
181469	& KIC 4150611	 	  &08.00 &A2	     & 7.73 (46) &  D	&    94.090 &	    &	 	 &	  & 7400 (100) &	    &3.80 (20)   &  128 (5)	  \\
	& KIC 4544587		  &10.83 &	     & 1.36 (41) &  D	&    2.1891 & 0.0208&		 &	  & 8600 (100) &7750 (180)  &4.12 (02)   &  87 (13)	  \\
	& KIC 4739791		  &14.63 &A7V	     &	 	 &  D	&    0.8989 & 0.0482&		 &	  & 7778 (28)  &5447 (17)   &4.20 (02)   &	   	  \\
	& KIC 6220497		  &	 &	     &	 	 &  SD  &    1.3232 & 0.1174&		 &	  & 7279 (54)  &3907 (22)   & 3.78 (30)  &	   	  \\
	& KIC 6629588		  &	 &	     &	 	 &  D	&    2.2645 & 0.0746&		 &	  & 6787 (247) &4405 (621)  &	 	 &	   	  \\
	& KIC 8569819		  &	 &	     &	 	 &  D	&    20.849 &	    &	 	 &	  & 7100 (250) &6047 (253)  &	 	 &	   	  \\	
	& KIC 9851944		  &11.42 &	     &	 	 &  D	&    2.1639 & 0.0962&		 &	  & 7026 (50)  &6950 (50)   & 3.96	 &   53 (7)	  \\
	& KIC 10619109		  &11.90 &	     &	 	 &  SD  &    2.0452 & 0.0234&		 &	  & 7138 (284) &3824 (571)  &	 	 &	   	  \\
	& KIC 10661783		  & 9.53 &A2	     & 1.94 (26) &  SD  &    1.2314 & 0.0355&		 &	  & 7764 (54)  &5980 (72)   &3.90	 &   79 (4)   \\
	& KIC 10686876		  &11.54 &F0V	     &	 	 &  D	&    2.6184 & 0.0476&		 &	  & 8167 (285) &6475 (817)  &	 	 &	   	  \\	
	& KIC 11175495		  &	 &	     &	 	 &  SD  &    2.1911 & 0.0155&		 &	  & 8293 (290) &6999 (790)  &	 	 &	   	  \\
	& KIC 11401845		  &	 &	     &	 	 &  D	&    2.2000 &	    &	 	 &	  & 7590       & 	    &	 	 &	   	  \\		
	& TYC 7053-566-1	  &11.51 &	     & 1.09 (23) &  SD  &    5.1042 & 0.0743&		 &	  & 7000 (200) &4304 (9)    &3.91 (02)   &	   	  \\
	& USNO-A2.0 1200-03937339 &14.53 &	     &	 	 &  SD  &    1.1796 & 0.0326&  5.1 (4)   &	  & 7250       &4320 (108)  &3.90 (10)   &	   	  \\

 \hline
\end{tabular}
 \end{scriptsize}
\end{table}
 \end{landscape}

\setcounter{table}{0}
\begin{landscape}

\begin{table}
 \begin{scriptsize}

  \caption{Continuation.}
  \begin{tabular}{llllllllllllllll}
  \hline
  HD     & Name 		   &$i$        & $q$          & $f$    &$M_{p}$   &$M_{s}$   &$R_{p}$   &$R_{s}$   &$L_{p}$     &$L_{s}$   &$M_{bol,p}$&$M_{bol,s}$&$a_{p}$ &$a_{s}$ 	  & References	\\
         &                         & ($^{0}$)  &              &        &($M_{\sun}$)&($M_{\sun}$)&($R_{\sun}$)&($M_{\sun}$)&($L_{\sun}$) &($L_{\sun}$) &(mag)     &(mag)    & ($R_{\sun}$) & ($R_{\sun}$) &	\\  
 \hline
354963	 &QY Aql   	           &88.6 (5)   &0.250 (200)   &0.403   &1.6 (2)   &0.4 (1)   &4.1 (2)   &5.4 (2)  &43.0 (3.0)  &8.0 (1)   &	    &	       &4.0 (2)    &16.3 (7)  	& 6, 42 	              \\
193740	 &XZ Aql  		   &84.8 (1)   &0.204 (2)     &0.500   &2.5 (1)   &0.5 (03)  &2.3 (04)  &2.5 (04) & 6.0 (04)   &  0.5 (07)&  1.1 (1)& 3.6 (2)  &   10.1 (1)&	   	&  6, 73	     	     \\
	 &V729 Aql	  	   &77.3 (2)   &0.440 (10)    &0.723   &1.5 (2)   &0.7 (1)   &2.0 (01)  &2.0 (01) & 7.8 (1)    &  1.3 (2) &  2.0 (1)&  4.6 (1) &	   &		&  43		     	   \\
	 &V1464 Aql	  	   &38.4 (2)   &0.710 (20)    &1.000   & 	  &2.1 (05)  &1.8 (01)  &         & 12.0 (3)   &  4.4 (02)&	    &	       &   4.8     &		&  6, 11, 84	     	  \\
  	 &CZ Aqr   		   &89.7 (1)   &0.490 (100)   &0.780   &2.0	  &1.0 (1)   &1.9 (1)   &1.8 (1)  &15.3 (9)    &2.9 (2)   &1.8 (6)  &3.6 (6)   &1.9 (2)    &3.8 (1) 	& 41		              \\
211705	 &DY Aqr   		   &75.4 (5)   &0.310 (200)   &0.489   &1.8 (2)   &0.6 (4)   &2.1 (1)   &2.7 (1)  &	       & 	  &	    &	       &9.4 (5)    &9.4 (5)  	& 2, 6  	              \\
203069	 &RY Aqr  		   &83.2 (4)   &0.204 (6)     &        &1.3 (1)   &0.3 (02)  &1.4 (07)  &1.9 (1)  & 56.0 (9)   &  1.4 (3) & 4.4 (2) & 2.8 (2)  &    7.6    &  		&  6, 55	     	    \\
	 &V551 Aur	           &74.3 (1)   &0.725 (6)     &0.539   & 	  &	     &	 	&	  &	       & 	  &	    &	       & 	   &		&  52		     	   \\
	 &EW boo   		   &76.5 (1)   &0.130 (2)     &0.708   &1.8 (2)   &0.2 (02)  &1.8 (1)   &1.1 (04) &10.9 (5)    &0.4 (1)   &	    &	       &5.0 (2)    &         	& 6, 83, 84	              \\
	 &YY Boo   		   &81.7 (1)   &0.290 (10)    &0.303   & 	  &	     &	 	&	  &	       & 	  &	    &	       & 	   &         	& 6, 24 	              \\
	 &Y Cam    		   &85.6 (1)   &0.241	      &0.535   &2.1 (1)   &0.5       &3.1 (05)  &3.3 (05) &1.4 (06)    &0.5 (06)  &1.3 (2)  &3.6 (2)   & 	   &	      	& 6, 25, 34, 41, 61, 84        \\
194168	 &TY Cap   		   &80.4 (2)   &0.520 (100)   &0.706   &2.0 (1.1) &1.1	     &2.5 (1)   &2.5 (1)  &24.3 (8)    &1.8 (2)   &1.3 (1)  &4.1 (1)   &2.7 (3)    &5.2 (1)    	& 6, 41 	              \\
	 &AB Cas   		   &88.3 (1)   &0.190	      &0.535   & 	  &	     &	 	&	  &	       & 	  &	    &	       & 	   &	      	& 6, 67, 84	              \\
	 &IV Cas   		   &87.5 (5)   &0.408 (1)     &0.781   &2.0 (1)   &0.8 (04)  &2.1 (04)  &1.8 (03) &1.3 (05)    &0.3 (05)  &1.4 (1)  &3.9 (1)   & 	   &	      	& 6, 30, 31	              \\
017138	 &RZ Cas   		   &82.0 (3)   &0.342 (1)     &0.494   &2.0 (2)   &0.7 (1)   &1.6 (1)	&1.9	  &	       & 	  &	    &	       &6.6 (3)    &          	& 6, 69, 74, 84                \\
	 &V389 Cas	  	   &81.8 (2)   &  	      &        &1.5 (01)  &1.5 (01)  &2.5 (02)  &2.6 (05) &	       &  1.6	  &  3.9 (1)& 0.1      &	   &		&  6, 37	     	    \\
	 &V1264 Cen	           &86.5 (1)   &0.220 (20)    &0.350   &1.5 (02)  &0.3 (02)  &2.4 (02)  &4.0 (01) & 6.7 (03)   &  3.9 (03)& 1.8     & 3.5      & 	   &		&  6, 8 	              \\	 
222217	 &XX Cep   		   &81.6 (1)   &0.173 (5)     &0.387   &2.5 (1)   &0.4 (01)  &2.3 (02)  &2.4 (02) &20.0 (3.0)  &2.1 (4)   &1.5 (2)  &3.9 (4)   &    	   &	       	& 6, 26, 36, 84              \\
	 &WY Cet   	           &81.8 (1)   &0.260 (10)    &0.514   &1.7	  &0.4 (01)  &2.2 (1)   &2.3 (1)  &14.0 (9)    &1.7 (1)   &1.9 (8)  &4.2 (8)   &1.8 (3)    &6.9 (1)     & 6, 41, 84	              \\
075747	 &RS Cha   	           &83.4 (3)   &0.709	      &        &1.9 (01)  &	     &1.9 (01)  &2.2 (06) &2.4 (06)    & 	  &	    &	       & 	   &	       	& 1, 3, 84	              \\
057167	 &R Cma    		   &81.7 (2)   &0.140	      &0.542   &1.7 (1)   &0.2 (1)   &1.8 (03)  &1.2 (07) &8.2 (2)     &0.5 (01)  &	    &	       &5.7	   &5.7      	& 4, 6, 56, 84                \\
	 &UW Cyg   		   &87.1 (1)   &0.140 (100)   &0.316   &1.9	  &0.3 (1)   &2.2 (1)   &2.9 (1)  &18.0 (9)    &2.6 (1)   &1.6 (4)  &3.7 (6)   &1.5 (2)    &11.2 (1)   	& 6, 41	              \\
	 &V346 Cyg   	           &	       &  	      &        &2.3	  &1.8	     &3.8       &4.7	  &61.8	       &39.9	  &	    &	       & 	   &	       	& 6, 68, 84	              \\
	 &V469 Cyg    		   &81.0       &0.430	      &        &3.3	  &	     &2.7	&	  &	       & 	  &	    &	       & 	   &	        & 5, 41, 68	              \\
099612	 &AK Crt		   &	       &  	      &        & 	  &	     &	 	&	  &	       & 	  &   	    &	       &	   &		& 6, 60	              \\	 
	 &BW Del       		   &78.6 (4)   &0.160 (20)    &0.416   &1.5 (2)   &0.3 (1)   &2.1 (1)   &2.2 (1)  &10.0 (1)    &1.2 (1)   &1.3 (1)  &8.0 (4)   & 	   &	        & 42		              \\
152028	 &GK Dra       		   &86.1 (2)   &1.244 (20)    &0.356   &1.5 (1)   &1.8 (1)   &2.4 (04)  &2.8 (05) &	       & 	  &2.0 (1)  &1.8 (1)   & 	   &	        & 6, 19, 84, 89              \\
	 &GQ Dra		   &	       &  	      &        & 	  &	     &	 	&	  &	       & 	  &   	    &	       &	   &		&  51		     	    \\
172022	 &HL Dra       		   &66.5 (1)   &0.370 (100)   &0.859   &2.5 (2)   &0.9 (1)   &2.5 (4)   &1.8 (3)  &24.3 (7)    &1.9 (1)   &1.3 (2)  &4.1 (2)   &1.7 (3)    &4.4 (1)     & 6, 41, 84	              \\
173977	 &HN Dra       		   &67.0       &0.931	      &        &1.9       &1.3       &2.9       &1.4	  &	       & 	  &	    &	       &	   &		& 6, 7, 84	              \\
187708	 &HZ Dra       		   &72.0 (3)   &0.120 (40)    &0.773   &3.0 (3)   &0.4 (1)   &2.3 (1)   &0.8 (1)  &45.0 (3.0)  &0.4 (2)   &0.6 (4)  &5.9 (4)   &0.6 (1)    &4.7 (1)	& 6, 41, 84	              \\
	 &OO Dra  		   &85.7 (1)   &0.097 (2)     &0.558   &2.0 (3)   &0.3 (03)  &2.0 (1)   &1.2 (05) &	       & 	  &	    &	       & 	   &		& 6, 86	              \\
238811	 &SX Dra       		   &85.3 (1)   &0.373 (2)     &0.320   &1.8	  &0.5       &2.3	&4.3	  &16.4        & 	  &	    &	       &	   &		& 6, 72	              \\
139319	 &TW Dra       		   &86.8 (3)   &0.411 (4)     &0.581   &2.2 (1)   &0.9 (05)  &2.6 (02)  &	  &	       & 	  &1.3 (1)  &3.2       &12.2 (2)   &		& 6, 33, 75, 84              \\
	 &TZ Dra       		   &77.6 (1)   &0.310 (30)    &0.665   &1.8 (2)   &0.6 (1)   &1.7 (1)   &1.5 (1)  &9.0 (1.0)   &1.3 (1)   &	    &	       &1.2 (1)    &4.0 (2)	& 6, 42	              \\
021985	 &AS Eri		   &	       &0.277	      &        &1.9	  &	     &1.6	&	  &	       & 	  &	    &	       & 	   &		& 6, 41, 58, 84              \\
	 &TZ Eri       		   &87.7 (07)  &0.177 (5)     &0.306   & 	  &	     &	 	&	  &	       & 	  &	    &	       & 	   &		& 6, 39	              \\
336759	 &BO Her  		   &85.4 (4)   &0.220 (200)   &0.324   &1.8 (2)   &0.4 (1)   &2.5 (1)   &3.8 (1)  & 20.0 (1.0) & 4.6 (4)  &	    &	       & 2.7 (1)   & 12.1 (5)	& 6, 42	              \\
	 &CT Her  		   &81.9 (01)  &0.141	      &0.432   &2.3 (02)  &0.3 (04)  &2.1 (06)  &1.9 (08) & 17.4 (2.4) & 1.2 (2)  & 1.7 (2) & 4.5 (2)  & 	   &		& 1,6, 44	              \\
	 &EF Her  		   &77.80      &0.210	      &0.421   & 	  &	     &	 	&	  &	       & 	  &	    &	       & 	   &		& 6, 64	              \\
151973	 &LT Her  		   &75.6 (2)   &0.200 (3)     &0.840   &2.5	  &0.5       &2.7	&1.6	  & 49.5       &  1.7	  &	    &	       & 	   &		&  6, 62	     	   \\	 
	 &TU Her		   &	       &  	      &        & 	  &	     &	 	&	  &	       & 	  &   	    &	       &	   &		& 6, 45	              \\
	 &V948 Her	  	   &84.4 (6)   &0.270 (30)    &0.574   &1.5 (2)   &0.4 (07)  &1.6 (1)   &0.7 (3)  & 6.0 (1.0)  & 0.2 (1)  & 2.9 (2) & 7.0 (1.0)& 1.3 (7)   & 4.9 (2)	& 6, 40, 41	              \\
	 &AI Hya  		   &89.9 (1)   &  	      &        &1.9	  &2.1       &2.1	&3.8	  &	       & 	  & 1.5     & 1.2      & 	   &		& 6,29, 61, 68               \\
078014	 &RX Hya		   &	       &  	      &        & 	  &	     &	 	&	  &	       & 	  &   	    &	       &	   &		& 6, 33, 84	              \\
	 &AU Lac  		   &83.0 (1)   &0.300 (10)    &0.490   &2.0	  &0.6 (1)   &1.8 (1)   &2.1 (1)  & 12.6 (7)   & 0.8 (1)  & 2.0 (6) & 5.0 (7)  & 1.7 (2)   & 5.7 (1)	& 6, 41, 58	              \\
	 &WY Leo		   &	       &  	      &        & 	  &2.3       &3.3	&	  &	       & 	  &	    &	       & 	   &		& 5, 6, 15	              \\
	 &Y Leo   		   &86.1 (2)   &0.324 (3)     &        &2.3	  &0.7       &1.9	&2.5	  &	       & 	  &	    &	       & 8.6	   &		& 6, 75, 76, 84              \\
033789	 &RR Lep  		   &80.5 (6)   &0.287 (21)    &0.802   &2.5 (3)   &0.7 (1)   &2.4 (1)   &1.5 (2)  & 15.6 (4)   & 1.0 (1)  & 0.8 (4) & 4.6 (8)  & 5.1 (1)   & 5.9 (1)	& 6, 16, 42	              \\
	 &CL Lyn  		   &78.7 (1)   &0.190 (20)    &0.606   &2.0	  &0.4       &2.5 (1)   &1.9 (1)  & 25.2 (9)   & 2.0 (7)  & 1.2 (9) & 4.0 (8)  & 1.3 (3)   & 6.6 (1)	& 6, 41, 84	              \\
198103	 &VY Mic		   &	       &  	      &        &2.4       &2.0	     &2.2	&4.4  	  & 26.0       & 14.0	  &	    &	       & 	   &		& 6, 60, 68	              \\
	 &V577 Oph		   &	       &0.939 (6)     &        & 	  &1.6       &	 	&	  &	       & 	  &	    &	       & 	   &		& 6, 9, 88	              \\
155002	 &V2365 Oph	  	   &87.4 (1)   &0.538 (3)     &0.346   &2.0 (02)  &	     &1.1 (01)  &2.2 (01) & 35.0 (4.0) & 1.3 (03) & 0.9 (1) & 4.4      & 17.5	   & 17.5	& 6, 27, 41, 84              \\
293808	 &FL Ori  		   &84.5       &0.900	      &0.549   &2.9       &1.9       &2.1       &2.2      & 18.6  3.2  & 	  &	    &	       &	   &		& 6, 41, 68, 83              \\
248406	 &FR Ori                   &83.2 (1)   &0.325 (2)     &0.735   &1.8	  &0.6       &1.8	&1.6	  &	       & 	  &	    &	       &   5.2     &		&  6, 83	     	   \\
252973	 &V392 Ori	  	   &79.8 (03)  &0.247 (1)     &0.951   &2.0 (2)   &0.5 (05)  &2.0 (07)  &1.3 (04) & 16.9 (8)   &  0.5 (02)&	    &	       &   3.6 (1) &		&  6, 85	     	    \\
	 &MX Pav  		   &77.0       &0.150	      &        & 	  &	     &	 	&	  &	       & 	  &	    &	       &	   &		& 5, 6, 60	              \\
	 &BG Peg  		   &83.2 (1)   &0.233 (3)     &0.582   &2.2	  &0.5       &2.0	&2.4	  &	       & 	  &	    &	       &  9.2	   &		& 70		              \\
275604	 &AB Per		   &	       &  	      &        & 	  &	     &	 	&	  &	       & 	  &   	    &	       &	   &		& 31, 32, 84	              \\
	 &IU Per  		   &78.8 (4)   &0.273 (50)    &0.762   &2.2	  &0.6       &2.0	&1.5      & 19.1    1.1&	  &	    &	       &  5.4	   & 5.4	& 6, 35, 83	              \\
	 &AO Ser  		   &87.0 (1)   &0.396 (82)    &0.846   &2.4	  &1.0       &1.8	&1.5	  & 17.5    0.8&  1.6	  & 5.0     &	       &  5.8	   &		&  6, 22	              \\
  
  \hline
\end{tabular}
 \end{scriptsize}
\end{table}
 \end{landscape}

 \setcounter{table}{0}
\begin{landscape}

\begin{table}
 \begin{scriptsize}

  \caption{Continuation.}
  \begin{tabular}{llllllllllllllll}
  \hline
  HD     & Name 		   &$i$        & $q$          & $f$    &$M_{p}$   &$M_{s}$   &$R_{p}$   &$R_{s}$   &$L_{p}$     &$L_{s}$   &$M_{bol,p}$&$M_{bol,s}$&$a_{p}$ &$a_{s}$      & References\\
         &                         & ($^{0}$)  &              &        &($M_{\sun}$)&($M_{\sun}$)&($R_{\sun}$)&($M_{\sun}$)&($L_{\sun}$) &($L_{\sun}$) &(mag)     &(mag)    & ($R_{\sun}$) & ($R_{\sun}$)  &  \\  
 \hline
 	 &UZ Sge                   &88.8 (1)   &0.140 (100)   &0.396   &2.1 (2)   &0.3 (2)   &1.9 (2)   &2.2 (2)  & 19.0 (4)   &  1.9 (4) & 1.6 (2) & 4.0 (1)  &   1.2 (8) &   8.6 (6)	&  40	     	    \\
	 &AC Tau		   &	       &  	      &        & 	  &	     &0.0	&	  &	       & 	  &   	    &	       &	   &		&  6, 16, 41	        	\\
	 &IZ Tel		   &	       &  	      &        & 	  &	     &0.0	&	  &	       & 	  &   	    &	       &	   &		&  6, 60, 81	              \\
12211	 &X Tri   		   &87.9 (1)   &0.599 (200)   &0.724   &2.1	  &1.3       &	 	&	  &	       & 	  &	    &	       & 	   &		&  6, 51, 84	     	    \\
 115268	 &IO Uma  		   &78.3 (1)   &0.135 (3)     &0.342   &2.1 (1)   &0.3 (02)  &3.0 (04)  &3.9 (05) & 1.5 (04)   &  0.7 (07)& 1.1 (1) & 3.1 (2)  &  17.6 (1) &	   	&  6, 71, 84	     	   \\
	 &VV Uma  	 	   &80.9 (03)  &0.337 (2)     &0.722   & 	  &	     &1.7	&1.4	  &	       & 	  &	    &	       &   4.9     &		&  6, 20, 46,84      	    \\
	 &AW Vel		   &	       &  	      &        & 	  &	     &	 	&	  &	       & 	  &   	    &	       &	   &		&  6, 57	     	    \\
	 &BF Vel  	           &86.2 (1)   &0.424 (19)    &0.814   &2.0 (2)   &0.8 (08)  &1.8 (01)  &1.5 (01) & 15.2 (4)   &  1.8 (02)& 1.8     & 4.1      &   4.7     &		&  6, 54	     	   \\
	 &CoRot 105906206 	   &81.4 (1)   &0.574 (8)     &0.720   &2.3 (04)  &1.3 (03)  &4.2 (02)  &1.3 (01) &	       & 	  &	    &	       &  15.3 (1) &	  	&  6, 10	     	    \\
172189	 &GSC 455-1084    	   &73.2 (6)   &0.960 (1)     &0.621   &1.8 (2)   &1.7 (2)   &4.0 (1)   &2.4 (07) & 52.4 (2.9) & 22.2 (1) &	    &	       & 20.3 (1)  &	  	& 9, 41	              \\
232486	 &GSC 3671-1094 	   &	       &  	      &        & 	  &	     &	 	&	  &	       & 	  &	    &          &	   &            & 6, 17, 41, 84              \\
	 &GSC 3889-202  	   &	       &  	      &        & 	  &	     &	 	&	  &	       & 	  &	    &          &	   &		& 6, 12	              \\
	 &GSC 4293-432  	   &	       &  	      &        & 	  &	     &	 	&	  &	       & 	  &	    &          &	   &		& 14		              \\
	 &GSC 4588-883    	   &78.5 (2)   &  	      &        & 	  &	     &	 	&	  &	       & 	  &	    &	       & 	   &		& 6, 13	              \\
062571	 &GSC 4843-2140   	   &73.0       &0.662 (16)    &        & 	  &	     &	 	&	  &	       & 	  &	    &	       & 	   &		& 28, 41	              \\
220687	 &GSC 5825-1038 	   &	       &  	      &        & 	  &	     &	 	&	  &	       & 	  &	    &          &	   &		& 6, 60	              \\
	 &KIC 3858884		   &	       &0.999 (5)     &0.213   &1.9 (03)  &1.9 (04)  &3.5 (01)  &3.1 (01) &	       & 	  &	    &	       &   57.2 (2)&	   	&  6, 53, 84	     	     \\
181469	 &KIC 4150611		   &	       &  	      &        & 	  &	     &	 	&	  &	       & 	  &	    &          &	   &		&  6, 65, 84	     	   \\
   	 &KIC 4544587	           &87.9 (03)  &0.810 (12)    &0.689   &2.0 (1)   &1.6 (06)  &1.8 (03)  &1.6 (03) &	       & 	  &	    &	       &   1.6     &		&  6, 23	     	   \\
	 &KIC 4739791	           &72.6 (02)  &0.070	      &0.740   &  1.8 (1) &0.1 (06)  &1.7 (03)  &0.9 (02) & 10.0 (1.0) &  0.6 (1) &  2.3 (1)& 5.2 (2)  & 	   &		&  48		     	    \\
	 &KIC 6220497	  	   &77.3 (3)   &0.243 (10)    &0.871   &1.6 (8)   &0.4 (2)   &2.7 (6)   &1.7 (4)  & 18.0 (2)   &  0.6 (1) & 1.6 (1) & 5.3 (2)  & 	   &		&  47		     	   \\
	 &KIC  6629588  	   &	       &  	      &        &1.2 (3)   &1.8 (7)   &	 	&	  &	       & 	  &	    &	       &	   &            &  51, 79	     	   \\
	 &KIC 8569819	  	   &89.9 (1)   &0.588	      &        &1.7	  &1.0       &	 	&	  &	       & 	  &	    &	       &   44.6    &		&  38		     	    \\
	 &KIC 9851944	  	   &74.5 (02)  &1.010 (30)    &0.432   &1.8 (1)   &1.8 (07)  &2.3 (03)  &3.2 (04) &	       & 	  &	    &	       &   10.7 (1)&   10.7 (1) &  21		     	  \\
	 &KIC 10619109  	   &	       &  	      &        &1.5 (3)   &2.1 (8)   &	 	&	  &	       & 	  &	    &	       &	   &		&  51, 79	     	     \\
         &KIC 10661783    	   &82.4 (2)   &0.091	      &0.744   &2.1 (03)  &0.2       &2.6 (02)  &1.1 (02) &	       & 	  & 1.4     & 4.3 (1)  & 	   &		& 6, 49, 66	              \\
	 &KIC 10686876  	   &	       &  	      &        &1.9 (2)   &2.4 (8)   &	 	&	  &	       & 	  &	    &	       &	   &	        &  51, 79	     	   \\
         &KIC 11401845  	   &	       &  	      &        & 	  &	     &	 	&	  &	       & 	  &	    &          &	   &		&  18		     	   \\
	 &KIC 11175495  	   &	       &  	      &        &2.0 (3)   &3.1 (5)   &	 	&	  &	       & 	  &	    &	       &	   &		&  51, 79	     	    \\
	 &TYC 7053-566-1  	   &71.13      &0.236 (4)     &0.337   &1.7 (1)   &0.4 (03)  &2.4 (07)  &4.2 (11) &	       & 	  &	    &	       & 	   &		&  6, 59	     	   \\
	 &USNO-A2.0 1200-03937339  &84.6 (2)   &0.190 (20)    &0.760   &1.6 (2)   &0.3 (1)   &2.2 (04)  &1.4 (03) & 12.4 (5)   &  0.7 (1) &  1.0 (1)&  5.0 (1) &	   &		&  43		     	   \\

   \hline
\end{tabular}
 \end{scriptsize}
 \begin{description}
  \item 1\,-\,\citet{2005A&A...442..993A}, 2\,-\,\citet{2014MNRAS.443.3022A}, 3\,-\,\citet{2008CoAst.157...47B}, 4\,-\,\citet{2011MNRAS.418.1764B}, 5\,-\,\citet{2004A&A...417..263B}, 6\,-\,\citet{2016arXiv160905175C}, 7\,-\,\citet{2004A&A...426..247C}
  8\,-\,\citet{2007MNRAS.382..239C}, 9\,-\,\citet{2010AN....331..952C}, 10\,-\,\citet{2009A&A...507..901C}, 11\,-\,\citet{2014A&A...565A..55D}, 12\,-\,\citet{2013PASA...30...16D}, 13\,-\,\citet{2008IBVS.5856....1D}, 14\,-\,\citet{2009IBVS.5883....1D}, 
  15\,-\,\citet{2009IBVS.5892....1D}, 16\,-\,\citet{2009CoAst.160...64D}, 17\,-\,\citet{2016NewA...48...33E}, 18\,-\,\citet{2005A&A...434.1063E}, 19\,-\,\citet{2014IAUS..301..413G}, 20\,-\,\citet{2003Obs...123..203G}, 21\,-\,\citet{2015IBVS.6148....1G}
  22\,-\,\citet{2016ApJ...826...69G}, 23\,-\,\citet{2015CoSka..45..106H}, 24\,-\,\citet{2013MNRAS.434..925H}, 25\,-\,\citet{2010IBVS.5949....1H}, 26\,-\,\citet{2015AJ....150..131H}, 27\,-\,\citet{2014NewA...27...95H}, 28\,-\,\citet{2008MNRAS.384..331I}
  29\,-\,\citet{2012phd}, 30\,-\,\citet{1989Ap&SS.155...53K}, 31\,-\,\citet{2010PASP..122.1311K}, 32\,-\,\citet{2006MmSAI..77..184K}, 33\,-\,\citet{2004ASPC..310..399K}, 34\,-\,\citet{2003A&A...405..231K}, 35\,-\,\citet{2002A&A...391..213K}, 36\,-\,\citet{2013BaltA..22..111K}, 
  37\,-\,\citet{2016AJ....151...77K}, 38\,-\,\citet{2015NewA...40...64K}, 39\,-\,\citet{2015MNRAS.446.1223K}, 40\,-\,\citet{2008CoAst.157..336L}, 41\,-\,\citet{2012NewA...17..634L}, 42\,-\,\citet{2012MNRAS.422.1250L}, 43\,-\,\citet{2013Ap&SS.343..123L}, 
  44\,-\,\citet{2014Ap&SS.353..559L}, 45\,-\,\citet{2004IBVS.5572....1L}, 46\,-\,\citet{2001MNRAS.325..617L}, 47\,-\,\citet{2001MNRAS.325..617L}, 48\,-\,\citet{2016MNRAS.460.4220L}, 49\,-\,\citet{2016AJ....151...25L}, 50\,-\,\citet{2013A&A...557A..79L}, 
  51\,-\,\citet{2017MNRAS.465.1181L}, 52\,-\,\citet{2010ASPC..435..101L}, 53\,-\,\citet{2012RAA....12..671L}, 54\,-\,\citet{2014A&A...563A..59M}, 55\,-\,\citet{2009Ap&SS.323..115M}, 56\,-\,\citet{2016AJ....152...26M}, 57\,-\,\citet{2000IBVS.4836....1M}, 
  58\,-\,\citet{2013JAVSO..41..182M}, 59\,-\,\citet{2013PASJ...65..105N}, 60\,-\,\citet{2016A&A...587A..54N}, 61\,-\,\citet{2007AcA....57...61P}, 62\,-\,\citet{1988AJ.....95..190P}, 63\,-\,\citet{2010MNRAS.408.2149R}, 64\,-\,\citet{1983A&AS...52..311R}, 
  65\,-\,\citet{2008CoAst.157..365S}, 66\,-\,\citet{2012MNRAS.422..738S}, 67\,-\,\citet{2011MNRAS.414.2413S}, 68\,-\,\citet{2003AJ....126.1933S}, 69\,-\,\citet{2006MNRAS.370.2013S}, 70\,-\,\citet{2007ASPC..370..344S}, 71\,-\,\citet{2011NewA...16...72S}, 
  72\,-\,\citet{2013MNRAS.432.3278S}, 73\,-\,\citet{2013AJ....145...87S}, 74\,-\,\citet{2016NewA...46...40S}, 75\,-\,\citet{2009A&A...504..991T}, 76\,-\,\citet{2010AJ....139.1327T}, 77\,-\,\citet{2008IBVS.5826....1T}, 78\,-\,\citet{2011Ap&SS.331..105T}, 
  79\,-\,\citet{2016ApJ...824...15V}, 80\,-\,\citet{2007A&A...474..653V}, 81\,-\,\citet{2014AJ....147...35Y}, 82\,-\,\citet{2011NewA...16..157Z}, 83\,-\,\citet{2008IAUS..252..429Z}, 84\,-\,\citet{2013ApJ...777...77Z}, 85\,-\,\citet{2015AJ...149...96Z}, 
  86\,-\,\citet{2015AJ....150...37Z}, 87\,-\,\citet{2014AJ....148..106Z}, 88\,-\,\citet{2001IBVS.5087....1Z}, 89\,-\,\citet{2003A&A...404..333Z}.
 \end{description}

\end{table}
 \end{landscape}

\bsp    
\label{lastpage}
\end{document}